\pdfoutput=1
\documentclass[UKenglish,texlive=2016,cernpreprint,biber,texmf]{atlasdoc}
\usepackage{atlaspackage}
\usepackage{rotating}
\usepackage{soul}
 
\usepackage{atlasbiblatex}
 
\usepackage{atlascontribute}
\usepackage{comment}
\usepackage{subfig}
\usepackage{atlasphysics}
 
\graphicspath{{logos/}{figures/}}
\usepackage{siunitx}
\usepackage{booktabs}
\usepackage{multirow}
\usepackage{comment}
 
\usepackage{ANA-IDET-2020-01-PAPER-defs}
 
\AtlasTitle{Measurements of sensor radiation damage in \\ the ATLAS inner detector using leakage currents}
 
\author{The ATLAS Collaboration}
 
\PreprintIdNumber{CERN-EP-2021-055}

\AtlasJournal{Journal of Instrumentation }
\AtlasJournalRef{JINST 16 (2021) P08025}
\AtlasDOI{10.1088/1748-0221/16/08/P08025}
 
\AtlasAbstract{
Non-ionizing energy loss causes bulk damage to the silicon sensors of the ATLAS pixel and strip detectors.   This damage has important implications for data-taking operations, charged-particle track reconstruction, detector simulations, and physics analysis.  This paper presents simulations and measurements of the leakage current in the ATLAS pixel detector and semiconductor tracker as a function of location in the detector and time, using data collected in Run 1 (2010--2012) and Run 2 (2015--2018) of the Large Hadron Collider.  The extracted fluence shows a much stronger $|z|$-dependence in the innermost layers than is seen in simulation.  Furthermore, the overall fluence on the second innermost layer is significantly higher than in simulation, with better agreement in layers at higher radii.  These measurements are important for validating the simulation models and can be used in part to justify safety factors for future detector designs and interventions.
}
\hypersetup{pdftitle={ATLAS document},pdfauthor={The ATLAS Collaboration}}
 
\begin{document}
 
\maketitle

\section{Introduction}
\label{sec:intro}

The ATLAS pixel and strip detectors are the subdetectors in closest proximity to the interaction point and are exposed to an unprecedented amount of radiation.  Monitoring and modelling the bulk radiation damage in the pixel and strip detector sensors is crucial for many aspects of the ATLAS experiment~\cite{Aad:2008zzm} including radiation protection, determination of operational conditions, offline data analysis, and upgrade design. Understanding the impact of radiation damage will help extend the lifetime and optimal use of the detectors.  Sensors designed for the high-luminosity phase of the Large Hadron Collider (LHC) will need to cope with about an order of magnitude more fluence than the present detector and thus investigations with the current detector  will provide valuable input to preparations for those data-taking conditions in the near future.
 
One of the best-characterized methods for monitoring silicon radiation damage is based on measuring the sensor leakage current.  This paper documents measurements of the sensor leakage current in the ATLAS pixel and strip detectors for the entire first and second runs of the LHC.   Models of thermal annealing combined with temperature and luminosity histories in data are used to extract the measured fluence across detector regions.  These data are compared with simulations of radiation damage that model particle production, propagation through the sensors, and bulk damage.   Thermal annealing models can be used to extract the fluence from data for comparison with simulations or annealing models can be combined with simulations to compare with measured leakage currents directly.  Complementary studies related to the detailed modelling of the sensor response to deposited charge from minimum-ionizing particles can be found in Ref.~\cite{Aaboud:2019wgd}.
 
This paper is organized as follows.  Section~\ref{sec:detector} introduces the ATLAS pixel and strip detectors and reviews the effects of radiation damage on their sensors.  Next, Section~\ref{sec:sim} describes how the complex radiation fields inside the ATLAS inner detector can be modelled and how the corresponding damage leads to changes in the sensor leakage current.  Technical details of the measurements for each subdetector are presented in Section~\ref{sec:measurements}.  All of the measurements are compared in Section~\ref{sec:comparisons}.  The paper ends with the conclusions and outlook in Section~\ref{sec:conclusions}.
 
\section{The ATLAS inner-detector silicon sensors and radiation damage effects}
\label{sec:detector}
 
\begin{figure}[!ht]
\centering
\includegraphics[width=0.8\textwidth]{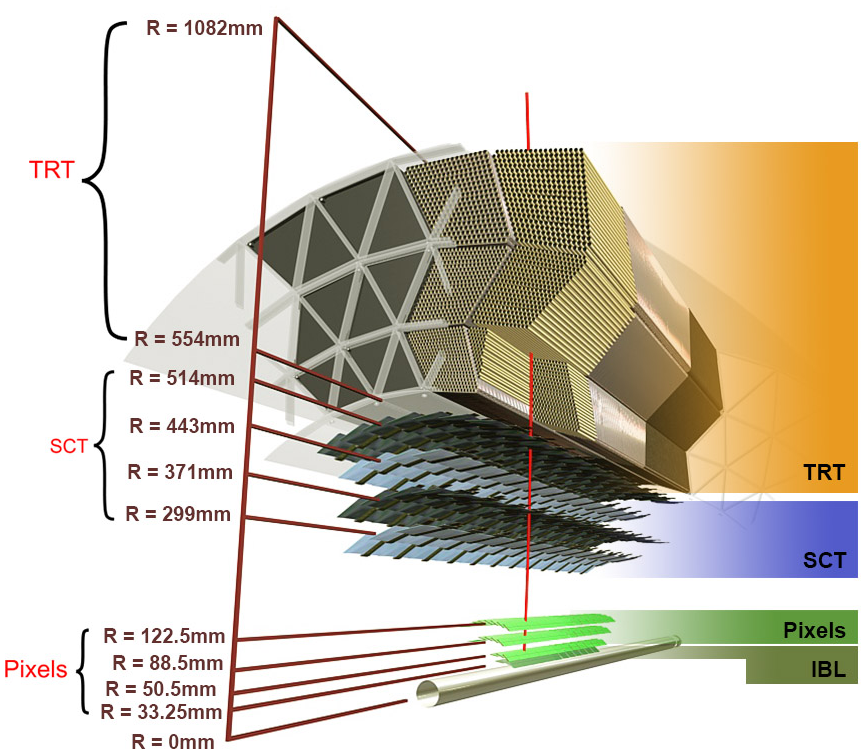}
\caption{A schematic view of the ATLAS inner detector. Radially outward from the collision point are the ATLAS insertable B-layer (IBL), the other layers of the pixel detector, the semiconductor microstrip tracker SCT, and the transition radiation tracker (TRT).  A red curved line represents a charged particle traversing the various layers and bending in the 2~T magnetic field.  The innermost pixel layer is called the Insertable $B$-Layer and was added to the detector between the first and second runs of the LHC.}
\label{fig:schematic}
\end{figure}

The ATLAS inner detector is composed of three subdetectors immersed in a 2~T magnetic field for measuring the trajectories of charged particles.  The two innermost subdetectors are based on silicon pixel and strip sensors, respectively.  Figure~\ref{fig:schematic} illustrates the radial extent of the pixel detector, the strip detector (called the semiconductor tracker or SCT) and transition radiation tracker (TRT).
 
The ATLAS pixel detector~\cite{Aad:2008zzm,Aad:2008zz,Capeans:1291633,Abbott:2018ikt} consists of four barrel layers and two identical endcap regions, each with three disk layers.
The layers are composed primarily of $n^+$-in-$n$ planar oxygenated~\cite{Lindstrom:421210,Lindstrom:2000gd} silicon sensors.
The four barrel layers, labelled Insertable $B$-Layer (IBL)~\cite{Capeans:1291633,Abbott:2018ikt}, $B$-Layer, Layer-1 and Layer-2, are arranged in concentric cylinders at radii of 33.25, 50.5, 88.5, and 122.5~mm from the beam axis.  While most of the pixel detector was installed before the start of LHC Run~1, the innermost barrel layer, IBL,  was installed during the shutdown (LS1) between LHC Run~1 and Run~2.  Most of the sensors in the pixel detector (IBL) are $250$ ($200$)~\textmu m  thick with a traditional planar geometry.  At high $|z|$ values\footnote{ATLAS uses a right-handed coordinate system with its origin at the nominal interaction
point (IP) in the centre of the detector and the $z$-axis coinciding with the axis of the beam pipe.  The
$x$-axis points from the IP towards the centre of the LHC ring, and the $y$-axis points upward.
Cylindrical coordinates ($r$,$\phi$) are used in the transverse plane, $\phi$ being the azimuthal angle
around the $z$-axis. The pseudorapidity is defined in terms of the polar angle $\theta$ as $\eta = - \ln
\tan(\theta/2)$. 
} the IBL contains $n^+$-in-$p$ 3D sensors~\cite{PARKER1997328} that are $230$~\textmu m  thick.   The pixel pitch is $50\times 250$~\textmu m$^2$ for the IBL and $50\times 400$~\textmu m$^2$ for the other pixel layers.  
 
The SCT~\cite{Aad:2008zzm,IDET-2013-01,Abdesselam:2006wt,Ahmad:2007zza,Abdesselam:2007ec} consists of four barrel layers and two endcaps, with nine disks each.  The four barrel layers are located at 29.9, 37.1, 44.3, and 51.4~cm from the centre of the ATLAS detector and the disks are located at $|z|$ values ranging from 74.9~cm to 272~cm in order to provide tracking coverage up to $|\eta|=2.5$.  All of the modules are composed of pairs of sensors offset by 40~mrad; this stereo angle provides space-point information.  The sensors have a pitch of 80~$\mu$m and are 285~$\mu$m thick.  Each sensor is constructed of high-resistivity $n$-type bulk silicon with $p$-type implants.  
 
Radiation damage in the sensor bulk is primarily caused by displacement of a silicon
atom from of its lattice site, resulting in a silicon interstitial site and a leftover vacancy (Frenkel pair)~\cite{Moll:1999kv,niel}.  These primary defects build, depending on the recoil energy, cluster defects and point defects in the silicon lattice that produce energy levels in the band gap.  When activated and occupied, these states increase the sensor leakage current ($I_\text{leak}$), which is proportional to the fluence received: $\Delta I_\text{leak}=\alpha\,\Phi_\text{eq} \,V$, where the effective fluence, $\Phi_\text{eq}$, is the 1~\MeV\ neutron equivalent fluence defined as the number of 1~\MeV\ neutrons applied to a sensor of surface area 1 cm$^2$ that cause damage equivalent to that from all particles that traversed the sensor.   The volume $V$ is the depleted volume of the silicon sensor and $\alpha$ is the current-related damage coefficient.  The goal of this measurement is to compare $I_\text{leak}$ with predictions of $\Phi_\text{eq}$, either by transforming the leakage current to a fluence or by transforming the fluence into a leakage current via $\alpha$.

\section{Simulations}
\label{sec:sim}
 
\subsection{Radiation simulations}
 
The complex radiation fields inside the ATLAS inner detector are simulated by propagating particles from inelastic proton--proton interactions, generated by \textsc{Pythia} 8~\cite{Sjostrand:2006za,Sjostrand:2014zea} using the MSTW2008LO parton distribution functions~\cite{Martin:2009iq} and the A3 set of tuned parameters~\cite{ATL-PHYS-PUB-2016-017}, through the ATLAS detector material using the particle transport code \textsc{Fluka}~\cite{Battistoni:2007zzb,Ferrari:898301} or \textsc{Geant}4~\cite{art:GEANT4}.  The particle and energy spectra are then folded with silicon damage factors from the RD50 database~\cite{273529,,HUHTINEN1993580,KONOBEYEV1992117,osti_10115441,art:Moll} to compute the 1~\MeV~neutron equivalent damage.  The tabulated weights cover neutrons, protons,
charged pions and electrons. For charged kaons the pion weights are used,
while for anti-neutrons, anti-protons, baryons and ions the proton weights are
used and positrons are treated like electrons.
 
The \textsc{Fluka} and \textsc{Geant}4 programs are composed of many subroutines based on a variety of phenomenological and first-principles models for nuclear and electromagnetic interactions of particles with the ATLAS detector material.  The most important difference from the point of view of particle propagation is the detector geometry, which slightly differs between the two programs as is explained in more detail below.
 
\subsubsection{\textsc{Fluka}}

The \textsc{Fluka} code~\cite{FLUKA1,FLUKA2} is well-established for studies of
hadronic and electromagnetic cascades induced by high-energy particles and is
the baseline code for radiation background simulations at CERN and the LHC
experiments.  Electrons, photons and muons up to $1000~\TeV$, and hadrons up to $20~\TeV$, can undergo interactions and be transported.
The lower cut-offs in energy for particle transport in the \textsc{Fluka} simulations are: hadrons and muons 100 keV; neutrons 10$^{-5}$~\eV\ (thermal); photons 30~\keV; and electrons 100~\keV.  However, photons and electrons have higher cut-offs in some regions (collimators, forward shielding) to reduce simulation time.  Anti-particles, heavy ions and
residual nuclei production are also treated by \textsc{Fluka}.

A complete description of \textsc{Fluka}'s physics models and capabilities can be found
in Ref.~\cite{FLUKA3} and references therein.  For example, inelastic hadron
interactions are described by different physics models depending on the
energy.  Inelastic hadron--hadron interactions above $5~\GeV$ are
treated by the Dual Parton Model~\cite{Cap94}, and below $5~\GeV$
by the resonance production and decay model~\cite{Fer96b}.  For hadron--nucleus
inelastic interactions above $5~\GeV$, Glauber--Gribov multiple
scattering followed by Generalized Intranuclear Cascade is employed.
Below $5~\GeV$ the pre-equilibrium cascade model \textsc{Peanut} is
used~\cite{Fer94,Fas95}.  All the above hadron interaction models include
evaporation and gamma de-excitation of the residual
nucleus~\cite{Fer96,Fer96a}.  Light residual nuclei are not evaporated, but are
fragmented into a maximum of six bodies according to a Fermi break-up model.

A description of the full ATLAS detector geometry and material has evolved in \textsc{Fluka} over the past 20~years,
and includes shielding, beam-line and machine components.
This has been developed independently of the ATLAS \textsc{Geant}4 geometry described below,
and in some cases has been simplified to speed up the simulations, e.g.\ by using cylinders to describe detector barrel layers.
In parts a three-dimensional geometry with $\phi$-asymmetry is implemented when considered important for providing more accurate predictions.
The magnetic fields are imported from the ATLAS offline software.

\subsubsection{\textsc{Geant}4}

\textsc{Geant}4~\cite{art:GEANT4} is used as the standard simulation toolkit for
physics analysis in ATLAS. The implementation of the detector geometry is
therefore very detailed, especially for instrumented regions and regions relevant
for the signal response, including upstream non-instrumented areas. The advent of new `physics lists' in \textsc{Geant}4 with high-precision transport of
neutrons and the possibility to simulate activation of nuclides and their
radioactive decay from timescales of nanoseconds to billions of years makes
\textsc{Geant}4 an attractive option for simulating the radiation background.
 
Various physics lists are used to determine the models and precision for processes simulated by \textsc{Geant}4.  Hadronic physics is governed by the  \texttt{FTFP\_BERT} list, which includes the Fritiof model~\cite{ANDERSSON1987289,Andersson:1996xi,NILSSONALMQVIST1987387,Ganhuyag:1997gz} with a precompound model above 4~\GeV\ and the Bertini intra-nuclear cascade model below 5~\GeV.   Energy thresholds are implemented via range cuts, whereby if the expected range of a secondary is less than some minimum value, the energy of that secondary particle is deposited at the end of the primary particle's step and no separate secondary is produced.  The range cuts vary from tens of microns to 1~mm depending on the subdetector material.
 
The geometry description for the ATLAS \textsc{Geant}4 simulation uses \textsc{GeoModel}~\cite{Boudreau:865601}, a library of basic geometrical shapes, to describe and construct the detector.  This model is the same one used for data analysis in ATLAS and is highly detailed for all detector components, including both active and passive material.  For the actual simulation, the geometry is translated entirely from the \textsc{GeoModel} to the \textsc{Geant}4 format.  Further details about the ATLAS \textsc{Geant}4 simulation can be found in Ref.~\cite{Aad:2010ah}.

\subsection{Modelling leakage current and annealing}
\label{eq:hamburg}
 
The formula $\Delta I_\text{leak}=\alpha\, V\, \Phi_\text{eq}$ only applies to instantaneous irradiation, where $\alpha$ is approximately independent of the damaging particles' energies and flavours.  After some time $t$ at a temperature $T$, the leakage current changes because of defect annealing, so $\alpha=\alpha(t,T)$.  Different models vary in their treatment of $\alpha$.  The model used to compare with all silicon layers is the Hamburg Model~\cite{Moll:1999kv}, as implemented in Ref.~\cite{Beyer2019}, where for $n$ time intervals, the predicated leakage current is given by
 
\begin{align}
\label{eq:leakage_modified}
I_\text{leak}=(\Phi/L_\text{int})\cdot\sum_{i=1}^n V_i\cdot L_\text{int,$i$}\cdot\left[\alpha_I\exp\left(-\sum_{j=i}^n\frac{t_j}{\tau(T_j)}\right)+\alpha_0^*-\beta\log\left(\sum_{j=i}^n\frac{\Theta(T_j)\cdot t_j }{ t_0}\right)\right],
\end{align}
 
where $L_\text{int,$i$}$ is the integrated luminosity, $t_i$ is the duration, and $T_i$ is the temperature in time interval $i$.  The first sum is over all time periods and the two sums inside the exponential and logarithm functions are over the time between the irradiation in time period $i$ and the present time.   The other symbols in Eq.~(\ref{eq:leakage_modified}) are $t_0=1\;$min, $V_i=\text{depleted volume}$ (in cm$^3$), $\alpha_I=(1.23\pm0.06)\times 10^{-17}\text{ A}/\text{cm}$, $\tau$ follows an Arrhenius equation $\tau^{-1}=(1.2^{+5.3}_{-1.0})\times 10^{13}\text{ s}^{-1}\times \text{e}^{(-1.11\pm 0.05)~\eV/k_\text{B}T}$, where $k_\text{B}$ is the Boltzmann constant, $\alpha_0^*=7.07\times 10^{-17}\;$A/cm, and $\beta=(3.29\pm 0.18)\times 10^{-18}$ A/cm.\footnote{A small temperature dependence has been observed in the value of $\beta$~\cite{Moll:1999kv}.  For this analysis, the reported value at 21\degC is used as it is closest to the operational temperature range of the detector.}   Note that the $\alpha$ and $\beta$ parameters are degenerate with the silicon damage factors, which are not well-known (see Sec.~\ref{sec:disc}). The temperature scaling function $\Theta(T)$ is defined by\footnote{This is not the only way to incorporate time-dependence in the thermal history.  Another proposal is to sum the inverse temperatures~\cite{Barney:1709387}.  Such a method has been compared with Eq.~(\ref{eq:tempscaling}) and results in similar predictions for the leakage current at the present fluence levels and annealing times.}
 
\begin{align}
\label{eq:tempscaling}
\Theta(T) = \exp \left[ - \frac{E_I^*}{k_\text{B} } \left( \frac{1}{T} - \frac{1}{T_\text{ref} } \right) \right],
\end{align}
 
where $E_I^*=(1.30\pm 0.14)$~\eV\ and $T_{\text{ref}}$ is a reference temperature.
 
The value of the fluence rate, $\Phi/L_\text{int}$, in data is estimated by performing a fit using Eq.~(\ref{eq:leakage_modified}) and letting only this parameter float.  The value $\Phi/L_\text{int}$ can be predicted from the \textsc{Pythia}+\textsc{Fluka} or \textsc{Geant}4 simulations.  When this value is used, the resulting predictions are called `unscaled'.  It is useful to fit a scale factor to this value instead of fitting $\Phi/L_\text{int}$ without any prior.  The simulation normalized with a scale factor obtained from a fit to data are defined as ``scaled'' in the following.
 
\section{Measurements}
\label{sec:measurements}
 
\subsection{General inputs and corrections}
 
\subsubsection{Luminosity}
 
Luminosity data are collected approximately once per minute.  The luminosity scale is determined by a set of dedicated bunch-by-bunch luminosity detectors~\cite{Aaboud:2016hhf} that are calibrated using the van der Meer method~\cite{vanderMeer:296752}. The absolute luminosity used in this study accounts for luminosities accumulated during times when the ATLAS detector is operating and also when it is not operating, because all particle fluence received by the silicon sensors will impact the leakage current. Luminosities used for this study surpass the quantities reported as the official ATLAS integrated luminosity usable for physics.
 
The uncertainty in the luminosity does not contribute to uncertainty in the leakage current data but is included in the fluence measurement.
The uncertainty in the combined 2015--2018 integrated luminosity is 1.7\%~\cite{ATLAS-CONF-2019-021}, obtained using the LUCID-2 detector~\cite{LUCID2} for the primary luminosity measurements.
 
A series of quality criteria are applied to the leakage current data considered in subsequent sections. Leakage current data are excluded for modules during periods when their bias voltage was not applied. Data collected within one minute of high-voltage turn-on are also excluded. Analysis of the leakage current data is restricted to times when the LHC has declared the proton beams to be stable. Luminosity recorded outside of these periods is included.
 
\subsubsection{Temperature corrections}
 
The leakage current depends on the temperature of the sensor~\cite{Moll:1999nh}. The following equation converts the leakage current of a sensor measured at temperature $T$ to that at a reference temperature $T_\text{R}$:
 
\begin{align*}
I_\text{leak}(T_\text{R})=I_\text{leak}(T)\,\left(\frac{T_\text{R}}{T}\right)^2\,\exp\left[-\frac{E_\text{eff}}{2k_\text{B}}\left(\frac{1}{T_\text{R}}-\frac{1}{T}\right)\right]\,,
\end{align*}
where $E_\text{eff}$ is the effective silicon band-gap energy after irradiation, also called the activation energy, and $k_\text{B}$ is the Boltzmann constant. A value of $1.21$~\eV\ was used for $E_\text{eff}$ for all sensors studied in Ref.~\cite{Chilingarov_tscale}. This choice provides consistency when comparing results from different subdetectors. A study using this value, performed with sensors that have been subjected to different radiation conditions, is presented in the next section.
 
\clearpage
\subsection{Optimal \Eeff study with the silicon sensors on the pixel layers and disks}
\label{sec:eff}
 
\begin{figure}[!ht]
\centering
\includegraphics[width=0.45\textwidth]{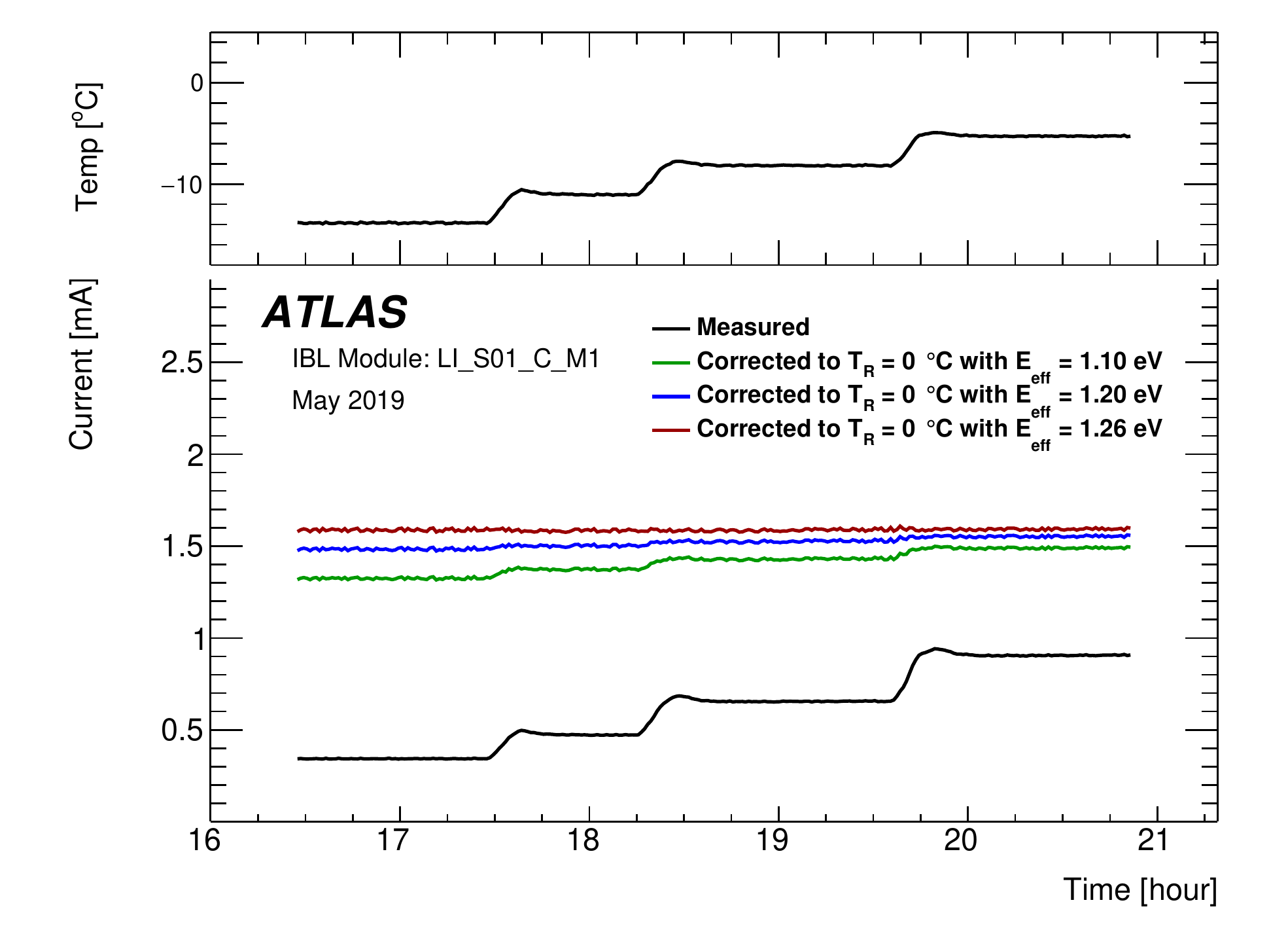}
\includegraphics[width=0.49\textwidth]{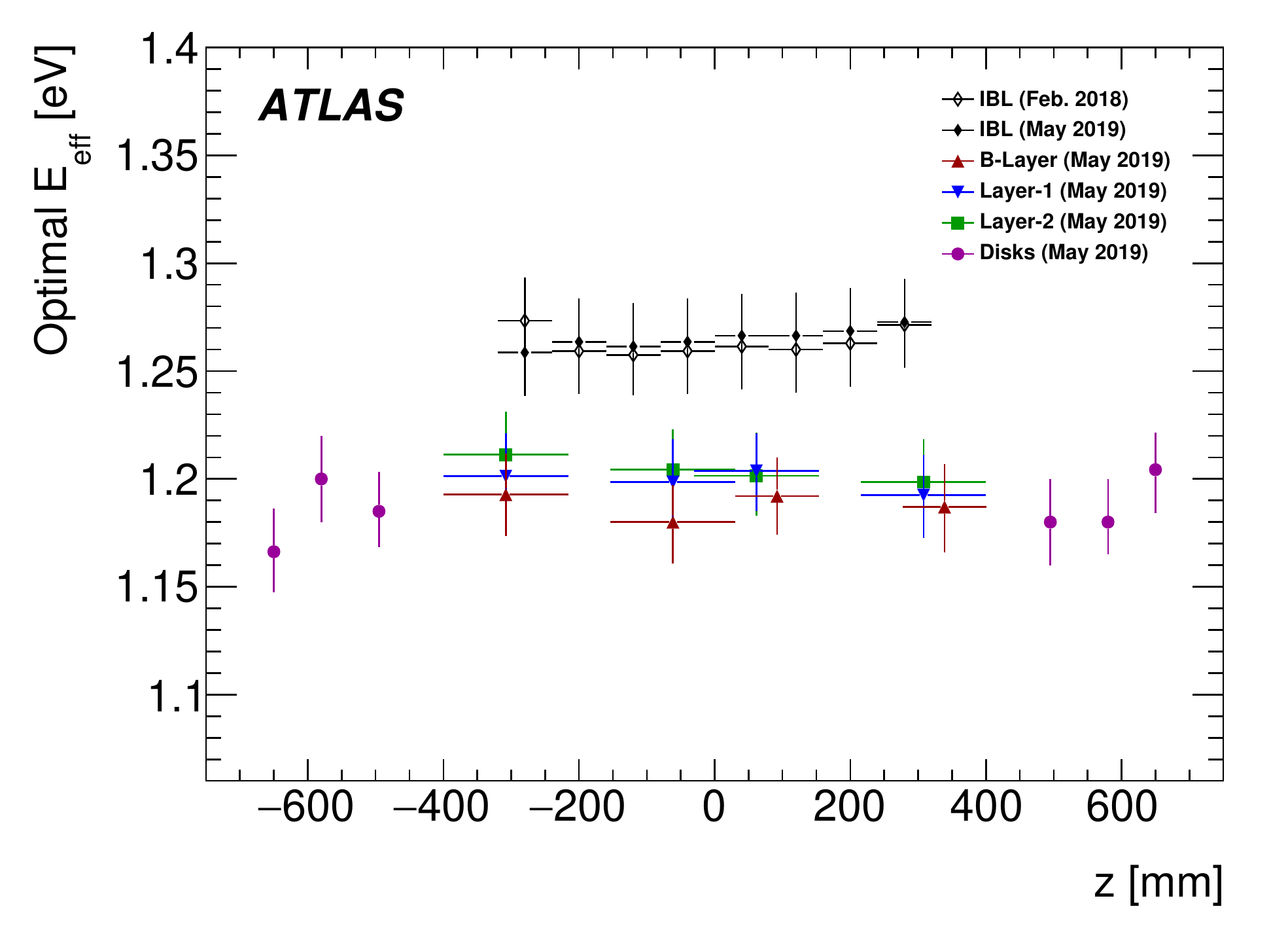}
\caption{Left: The best value of the effective silicon band-gap energy $(E_\text{eff}$) for use
in normalizing silicon sensor leakage current to a temperature other than that
at which it was recorded is investigated, for one module on the IBL.  The top panel shows the temperature of the pixel detector module as set to several fixed values,
and measured with the module temperature sensor.  The lower panel shows the leakage current data as measured (black line) with a clear temperature dependence.   Right: The optimal $E_\text{eff}$ value is determined for each module and then the average value is computed in bins of $z$ for each layer and disk.  The vertical error bars represent the impact on the optimal $E_\text{eff}$ value due to a $\pm 2$\degC conservative uncertainty in the module temperature.  Variations larger than this would be inconsistent with thermal models.  Each bin is defined according to the average position of the modules whose data are used. }
\label{fig:Eeffsummary}
\end{figure}

Dedicated temperature scans were used to measure \Eeff in the pixel detector.  The upper panel of the left plot in Figure~\ref{fig:Eeffsummary} shows the measured temperature as a function of time during the scan for a representative module in the IBL.   The bottom panel shows the measured leakage current and corrections to a reference temperature of 0\degC with several values of $E_\text{eff}$. The optimal value of $E_\text{eff}$ in the temperature correction equation is the value that results in corrected leakage current data that best fit a line of zero slope.  The best fit is determined using a minimum-$\chi^2$ figure of merit. The optimal $E_\text{eff}$ is 1.26~\eV\ for this particular module and corresponds to an integrated luminosity of 161 fb$^{-1}$ delivered to the IBL.   This procedure is repeated for all modules in the pixel detector and the extracted values of \Eeff are shown in the right plot of Figure~\ref{fig:Eeffsummary}.  The \Eeff values for the IBL modules were extracted using temperature scan data in Feb.~2018 (May~2019) corresponding to 95~(161)~fb$^{-1}$. The \Eeff values for all other layers correspond to the temperature scan data in May~2019 and 191~fb$^{-1}$ accumulated in Run~1 and Run~2. The bin ranges of the outer layers are determined by the paired-module powering scheme of module power supplies.
 
The measured \Eeff is found to vary by layer, with a small radial dependence.  The Layer-2 values are most consistent with $1.21$~\eV, with slightly lower values observed for Layer-1 and the $B$-Layer.  Higher values of \Eeff are observed for the IBL.  The data are consistent with a small ($\mathcal{O}(1\%)$) increase in \Eeff between 2018 and 2019 for the IBL.  It is possible that \Eeff depends on the composition of irradiating particles, the thermal history, or the sensor doping properties.   Further investigations are left to future studies.
 
\label{sec:Eeff}
 
\begin{figure}[!ht]
\centering
\includegraphics[width=0.49\textwidth]{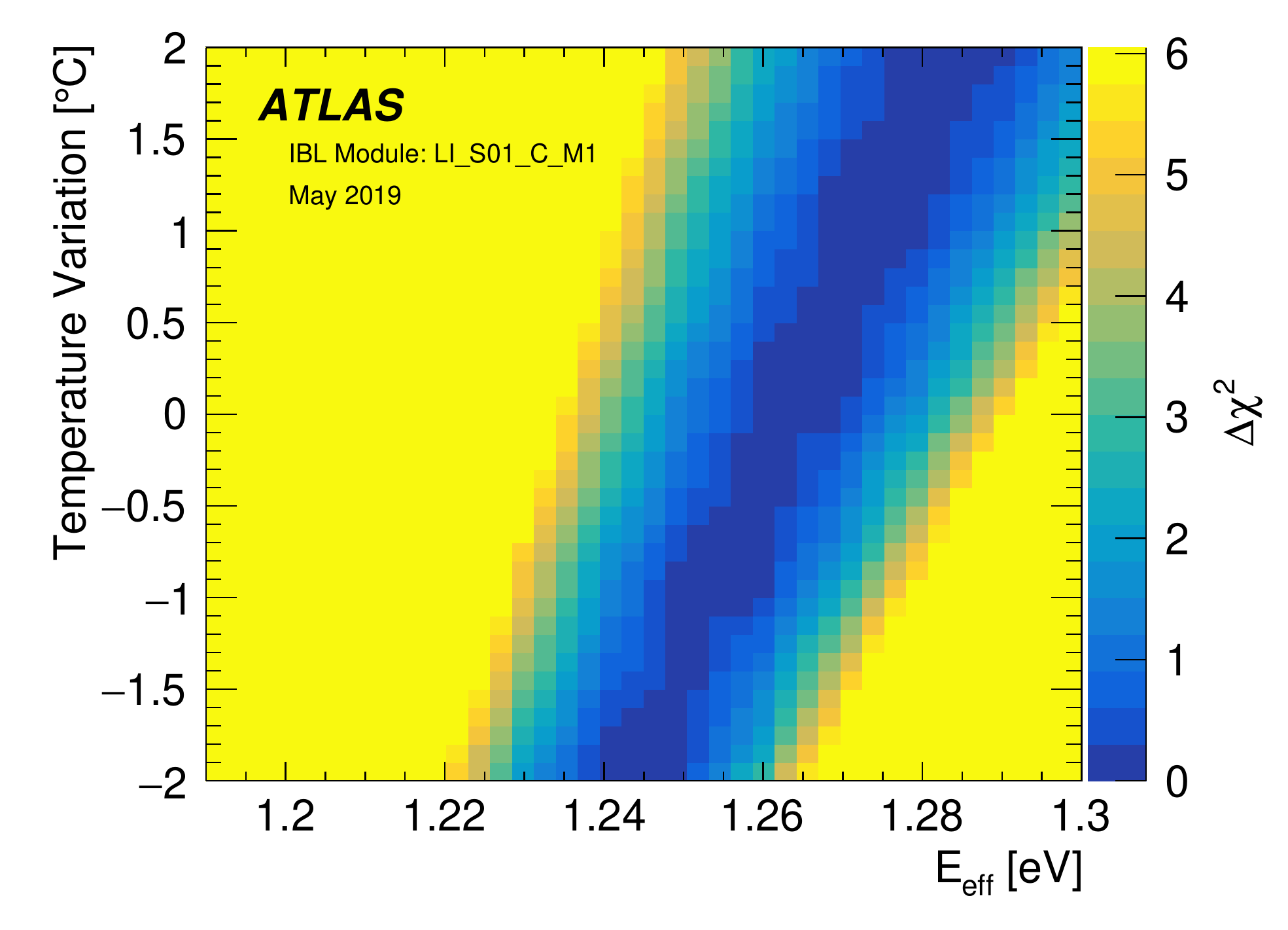}
\caption{The $\chi^2$ figure of merit is determined for a range of $E_\text{eff}$ values and variations of the module temperature data, for one module on the IBL. Steps of 0.01~\eV, in the range 0.5~\eV\ to 1.5~\eV, for $E_\text{eff}$ (a reduced range is shown for this model) and steps of 0.1\degC, in the range $-2.0$\degC to 2.0\degC, for the temperature variation are investigated independently. Here the definition of variation is the measured module temperature plus or minus a constant that corresponds to a systematic uncertainty in the temperature sensor reading. The measurement corresponds to data of integrated luminosity 161~fb$^{-1}$ delivered to the IBL. }
\label{fig:EeffTempChi2}
\end{figure}

The largest uncertainty in the \Eeff measurement is due to the offset between the measured and true sensor temperatures.  This offset is not well-constrained; a $\pm 2$\degC uncertainty is used for illustration purposes, acting as a conservative estimate to show the effects of such a shift on the final measurements.  In comparison, the statistical uncertainty is found to be negligible. In principle, one could simultaneously extract the temperature offset and \Eeff from the fit.  The $\Delta\chi^2$ landscape is presented in Figure~\ref{fig:EeffTempChi2} and shows that there is a near degeneracy between the extracted temperature offset and \Eeff.  Therefore, additional studies of the temperature offset using simulations or laboratory tests will be required to determine \Eeff with more precision.
 
\subsection{Innermost pixel layer (IBL)}
 
Each sensor is equipped with a negative temperature coefficient (NTC) thermistor for measuring the temperature.   On the FEI4 readout chips~\cite{GarciaSciveres:2011zz} bonded to the sensors, there is a 10-bit analog-to-digital converter associated with an 8-to-1 analogue multiplexer that can be used to select and read out the temperature, power supply voltages, voltage references, detector leakage current, and other detector control system analogue voltages.   The leakage currents were measured at the nominal operational temperature and bias voltage settings.  In particular, the temperature started at -2\degC and the high voltage for the planar (3D) sensors started at $-80$~V ($-20$~V) at the beginning of 2015 and was subsequently increased to ensure full depletion. The high voltage was increased to $-150$~V ($-20$~V) for the planar (3D) sensors at the end of 2016 corresponding to about 35~fb$^{-1}$, to $-350$~V ($-40$~V) in mid-2017 at about 45~fb$^{-1}$, and then to $-400$~V ($-40$~V) in 2018 at about 95~fb$^{-1}$.  The temperature changed to 10\degC during the first HV increase and then to -13\degC during the second HV increase.
 
During shutdown periods (such as year-end technical stops), the modules were powered off and no on-sensor temperature reading was available.  The cooling during these periods is determined with measurements from the nearby cooling pipes.
 
The uncertainty in the extracted fluence is dominated by a conservative 10\% uncertainty that accounts for the possible difference between the leakage current at the operational bias voltage and the current at the full depletion voltage. After irradiation, there is also an increase in leakage current with increasing bias voltage after full depletion, whereas the Hamburg Model predicts a constant leakage current above the full depletion voltage. Therefore, the choice of voltage for the leakage current measurement is important for comparison with the Hamburg Model prediction. Uncertainties due to the annealing model (0.1\%) and data fit (0.5\%) are subdominant.
 
\subsubsection{Results}
 
\begin{figure}[!ht]
\centering
\includegraphics[width=0.8\textwidth]{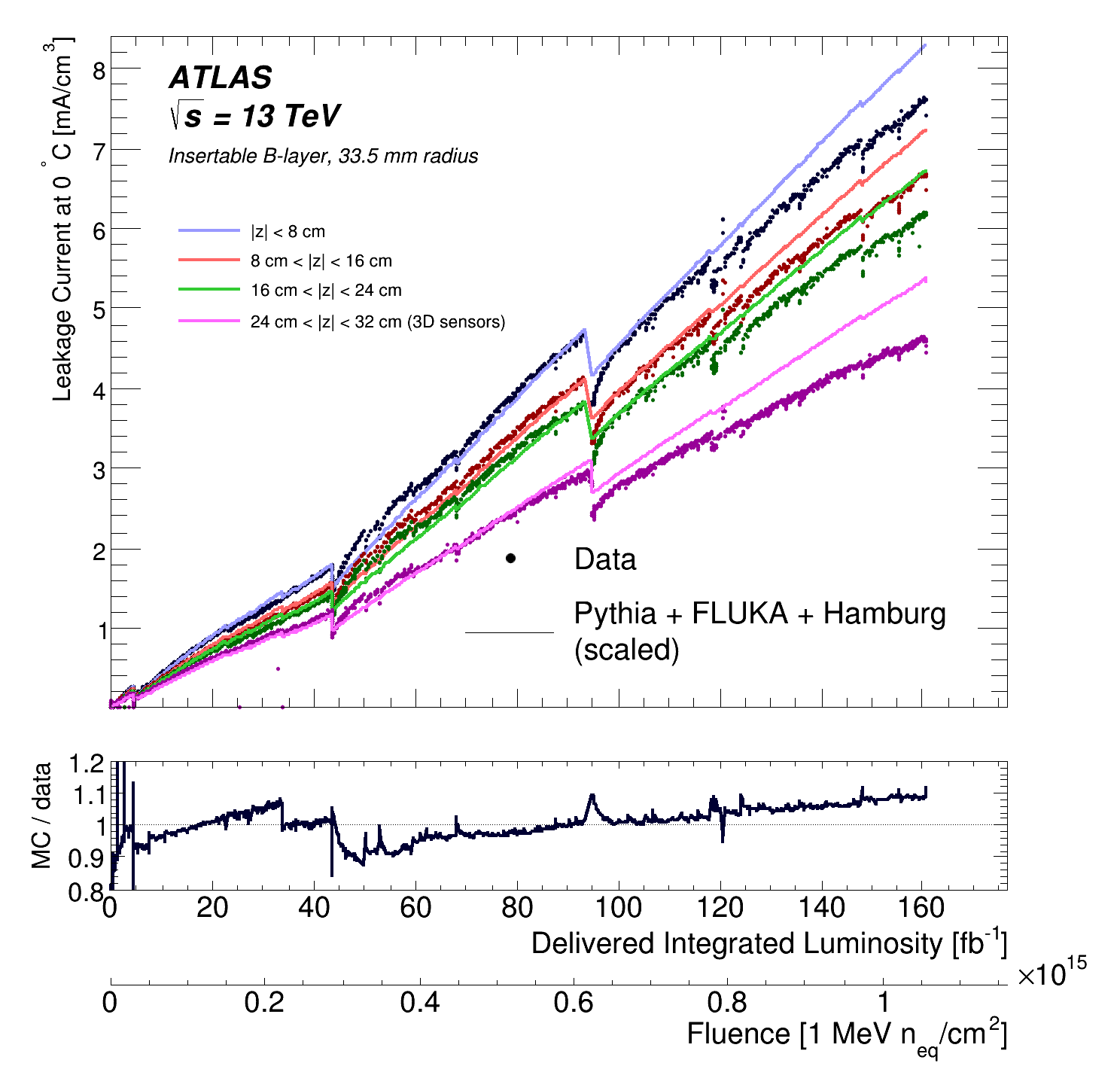}
\caption{The measured and predicted leakage currents for sensors on the Insertable $B$-layer, both normalized to 0\degC for four module groups spanning $|z|<8$~cm, $8<|z|<16$~cm, $16<|z|<24$~cm, and $24<|z|<32$~cm. Modules in the highest $|z|$ region use 3D sensors. The A and C sides of the detector ($z>0$ and $z<0$) are consistent with each other and averaged.  The dominant time-independent uncertainty of 10\% is not included to avoid overlapping bands.  The prediction is based on the thermal history of the modules combined with the Hamburg Model for modelling changes in the leakage current and \textsc{Pythia} + \textsc{Fluka} for simulating the overall fluence. For all four predictions, the overall scale normalization is based on a fit to the data across the entire range. Normalization factors are determined per $|z|$ region. The lower panel shows the ratio of the prediction to the data for the innermost module group. Similar MC/data trends are observed for the other three $|z|$ regions.  For illustration, the fluence is shown as a lower horizontal axis using the nominal luminosity-to-fluence from simulation at $|z|=0$.}
\label{fig:IBLTime}
\end{figure}

Measurements and scaled predictions averaged over the azimuthal angle $\phi$ and within module groups are presented in Figure~\ref{fig:IBLTime}.  The large drops in the current correspond to periods of annealing during year-end technical stops.  Apart from an overall constant offset, the model predictions agree with the data within about 10\% across Run~2.  Despite this good agreement, past 40~fb$^{-1}$ there is a clear monotonic trend in the ratio of the prediction to the data.  Judging from the structure in the data and prediction at the start of 2018 (around 95~fb$^{-1}$), the effects of annealing seem more pronounced in data than in the predictions.
 
\begin{figure}[!ht]
\centering
\includegraphics[width=0.8\textwidth]{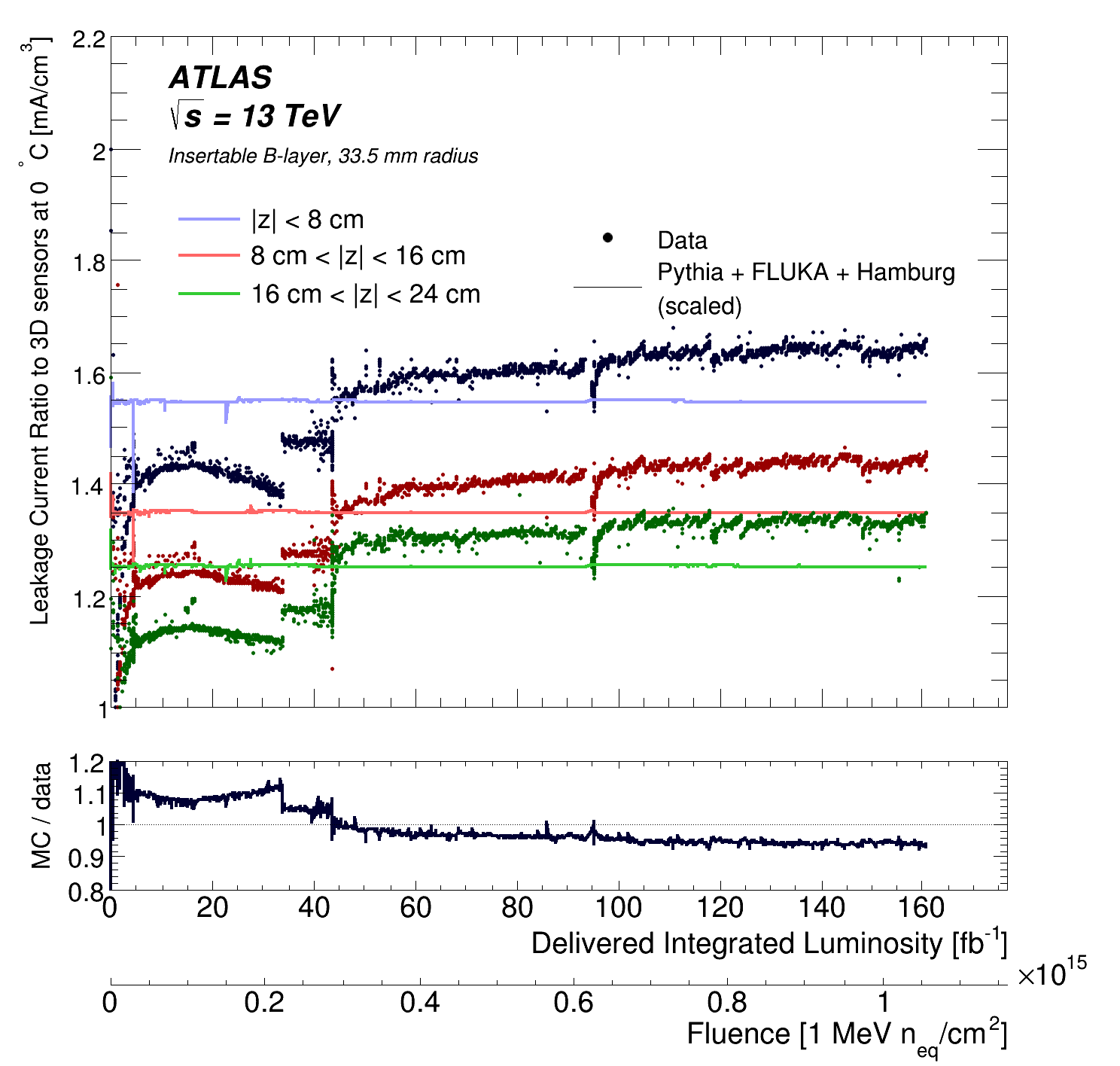}
\caption{The measured and predicted leakage currents on the Insertable $B$-layer, both normalized to 0\degC, divided by the currents in the 3D modules at $24<|z|<32$~cm. The other three module groups represent $|z|<8$~cm, $8<|z|<16$~cm, and $16<|z|<24$~cm. The A and C sides of the detector ($z>0$ and $z<0$) are consistent with each other and averaged.  The dominant time-independent uncertainty of 10\% is not included to avoid overlapping bands.  The prediction is based on the thermal history of the modules combined with the Hamburg Model for modelling changes in the leakage current and \textsc{Pythia} + \textsc{Fluka} for simulating the overall fluence. For all four predictions, the overall scale normalization is based on a fit to the data across the entire range. Normalization factors are determined per $|z|$ region. The lower panel shows the ratio of the prediction to the data for the innermost module group. Similar MC/data trends are observed for the other three $|z|$ regions.  For illustration, the fluence is shown as a lower horizontal axis using the nominal luminosity-to-fluence from simulation at $|z|=0$.}
\label{fig:IBLTimeRatio}
\end{figure}

Prior to 40~fb$^{-1}$, the planar sensors were slightly below full depletion.  This can be seen most clearly in Figure~\ref{fig:IBLTimeRatio}, which presents ratios of the leakage current in the planar sensors to the leakage current in the 3D sensors on the IBL.  Ratios of leakage currents are proportional to the ratio of the corresponding predicted fluences if the sensors are fully depleted.  The ratio is relatively constant past 40~fb$^{-1}$, but there is a clear decrease in the ratio between 15 and 35~fb$^{-1}$.  The 3D sensors require a much lower high voltage to be fully depleted and so assuming they are fully depleted during this period, the upper panel of Figure~\ref{fig:IBLTimeRatio} displays a measure of the depleted volume.\footnote{The ratio is predicted to be proportional to the depleted volume up to corrections that are induced by overdepletion.}  Once the high voltage was increased from $-80$~V to $-150$~V, the ratio of the planar sensor to 3D sensor leakage currents stabilized.
 
\subsection{Outer pixel layers and disks}
 
\subsubsection{Measurement subsystems}
 
Leakage current data are collected using two independent subsystems, the per-module high-voltage patch panel subsystem (HVPP4) and the multi-module power supply subsystem used to confirm and augment the HVPP4 measurement. The HVPP4 serves as a fan-out point for the bias voltages delivered to the pixel modules from the Iseg high-voltage power supplies~\cite{iseg} and monitors leakage currents at the pixel-module granularity level by means of a Current Monitoring Board system. Further details of the HVPP4 system can be found in Ref.~\cite{ATL-INDET-PUB-2014-004}.
 
During Run~1, each power supply channel was used to power and read out six or seven modules in parallel. During the long shutdown of the LHC between 2013 and 2015, more power supply units were installed for use in the pixel detector. During LHC Run~2, each power supply channel supplied one or two modules. The sum of the leakage current values is measured for the modules supplied by a common power supply unit. The precision of measurements of the leakage current read with the power supply units was improved during LHC Run~2. The raw leakage current data measured by both the HVPP4 system and the power supplies are stored in the COOL database~\cite{Valassi2008}.
 
Leakage currents were measured for each module individually using the HVPP4 subsystem in LHC Run~1 and Run~2 and for pairs of modules using the power supply subsystem in LHC Run~2. These measurements are subsequently averaged for each barrel layer and each disk. Averaged binned measurements are also examined for $\phi$ and $z$ dependence for a more refined investigation of the spatial dependence of the radiation damage.
 
\subsubsection{Precision and systematic uncertainties}
\label{sec:outeruncerts}
 
This section provides the elements that contribute to the uncertainty of the measurement of the leakage current, the Hamburg Model predictions, and the measurement of the fluence. The two subsystems, HVPP4 and the power supplies, are used to make independent measurements; the final uncertainty for each subsystem is calculated by adding all contributing elements (described below) in quadrature. The statistical uncertainty is used in the fit, over the full temporal range of the measurement, to determine a scale factor needed to bring the Hamburg Model into agreement with the data. The precision of current measurements made with the HVPP4 current monitoring circuit contributes a 12\% uncertainty. The precision of current measurements made with the power supply units contributes 4\% to the total measurement uncertainty~\cite{iseg}. The current measurements are made approximately once per minute and that interval contributes a 0.5\% uncertainty; this uncertainty is calculated by investigating the changes in the data over 10-minute intervals -- a time interval over which the leakage current is not expected to change. The precision of the temperature measurements contributes a 2.9\% uncertainty; the temperature is also not expected to fluctuate over short time intervals and is thus calculated in the same way as the current uncertainty. Uncertainties due to a possible difference between the temperature of the point on the module at which the temperature is measured and the point on the silicon sensor to which the temperature is attributed are found to be 10\% through changes to the modelled leakage current when a difference in temperature of 1\degC is applied. The total uncertainty of the leakage current data collected with HVPP4 is determined by adding in quadrature the HVPP4 current monitoring circuit measurement uncertainty, the current measurement uncertainty, and each of the two temperature uncertainties; the HVPP4 data uncertainty is found to be 15.9\%. The total uncertainty of the leakage current data collected with the dual-module power supply units is determined by adding in quadrature the power supply precision uncertainty, the current measurement uncertainty, and each of the two temperature uncertainties; the power supply data uncertainty is found to be 11.2\%. 
 
\subsubsection{Results}
 
\begin{figure}[!ht]
\centering
\includegraphics[width=0.9\textwidth]{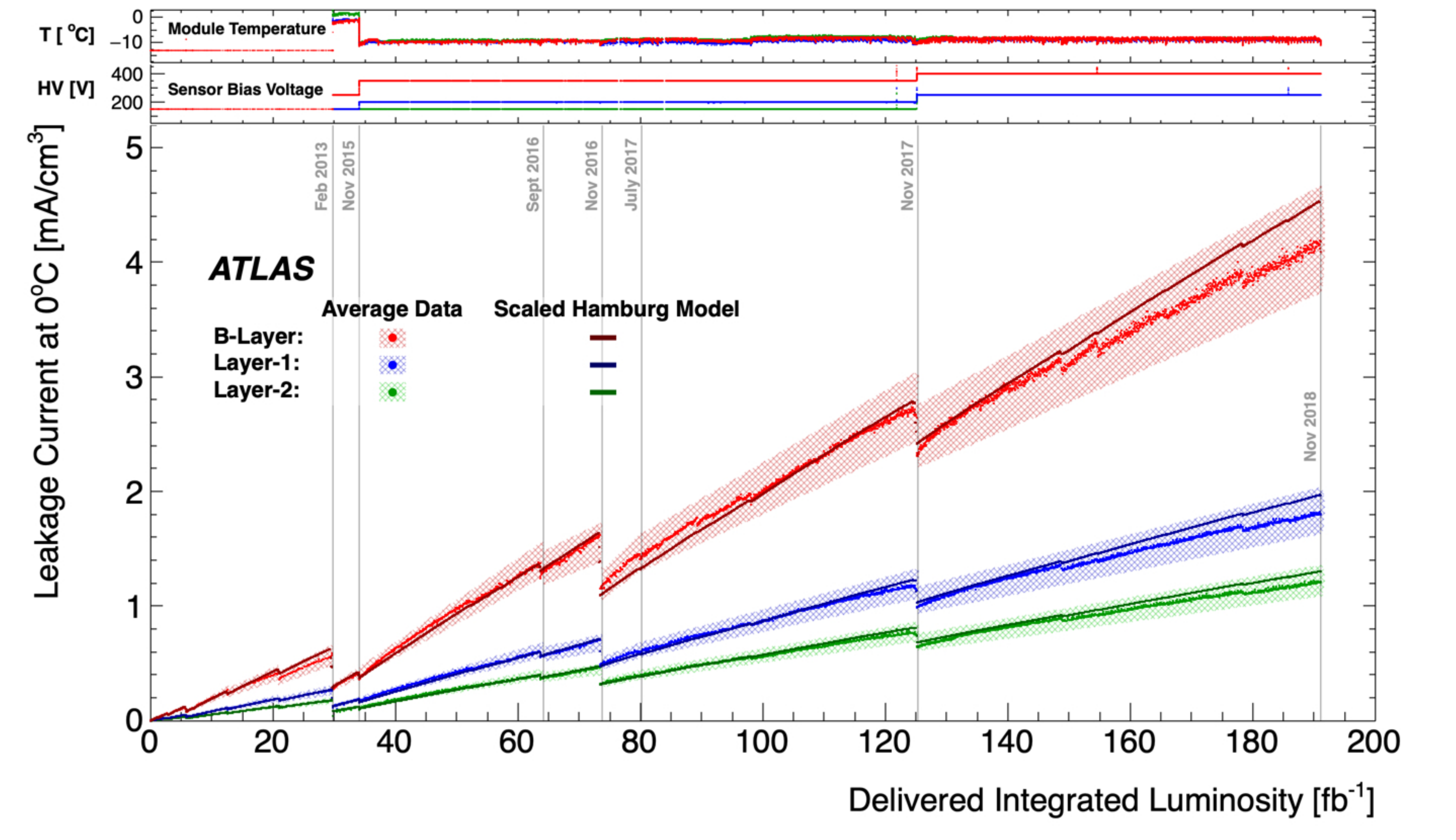}
\caption{Average measured leakage current of a representative sample of modules on the $B$-Layer, Layer-1 and Layer-2 over the full period of operation. The scaled prediction from the Hamburg Model is also shown.  The bands include uncertainties on the measurement, as described in Sec.~\ref{sec:outeruncerts}.}
\label{fig:OuterPixelTime}
\end{figure}

Measurements of leakage currents in all three barrel layers from February 2011 (early in LHC operation) to November 2018 are shown in  Figure~\ref{fig:OuterPixelTime}. The $B$-Layer (red points) shows the highest leakage current consistently. Layer-1 (blue points) and Layer-2 (green points) have progressively lower levels. Measurements on each layer are averaged over a representative sample of modules in $z$ and $\phi$. The measurements are consistent with expected higher levels of radiation for sensors closer to the interaction point. The Hamburg Model predictions have been scaled to match the measured leakage current with a luminosity-to-fluence scaling factor applied for each layer. The scale factors are determined in four $z$-binned regions for each layer.  The average of the scaled Hamburg Model predictions in each is used in a comparison with the average leakage current data. After the application of the scale factors, the Hamburg Model predictions fit well to the data throughout the full period of the measurement. The leakage current data are normalized to 0\degC; the average module temperature is shown in the top panel. Leakage current data are shown for periods of operation when the high voltage is applied across the silicon sensor; the average module bias voltage is shown in the middle panel of Figure~\ref{fig:OuterPixelTime}. Some spikes in the bias voltage are visible starting near the end of 2017, corresponding to bias voltage scans. Bias voltage scans prior to 2017 are not shown in the figure. Some dates corresponding to extended periods when the LHC beams were off, resulting in annealing of the sensors, are displayed within the lower panel with grey vertical lines. Not all such periods are marked. The module temperatures are taken to be 18\degC during these shutdown periods. During part of LS1, from February 2013 to February 2014 (LS1 ended in April 2015), the pixel detector was removed from the ATLAS cavern and kept at 22\degC.
 
The slopes of the leakage current plots in Figure~\ref{fig:OuterPixelTime} is due to constant damage from the applied fluence, and a slight difference is observed between data and prediction near the end of Run~2.    The drops in the leakage current are due to annealing; the annealing can also change the overall shape as a function of integrated luminosity. The observed differences are accounted for in the uncertainty bands, which are dominated by a temperature bias.  Other sources of uncertainty such as the luminosity uncertainty ($\mathcal{O}(1\%)$) are subdominant.
 
\begin{figure}[!ht]
\centering
\includegraphics[width=0.9\textwidth]{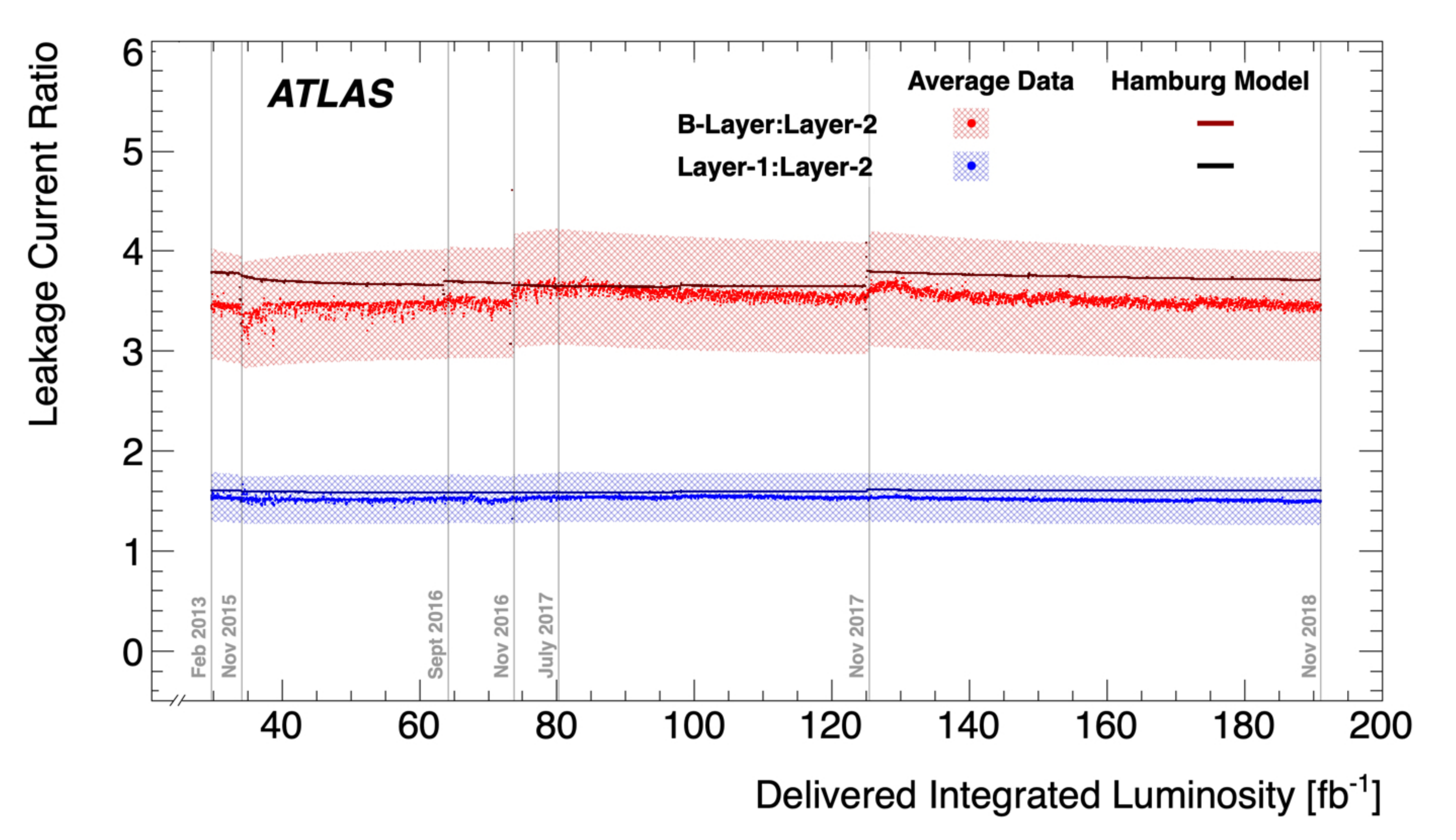}
\caption{Ratios of the $B$-Layer and Layer-1 leakage current data to Layer-2 leakage current for the LHC Run~2 period of ATLAS operation.  The bands include uncertainties on the measurement, as described in Sec.~\ref{sec:outeruncerts}}
\label{fig:OuterPixelTimeRatio}
\end{figure}

To reveal the impact of constant damage, a study of the ratios of the leakage currents across the layers is performed. Measured ratios of the average leakage current for modules on the $B$-Layer to the average leakage current for modules on Layer-2, and of the average leakage current for modules on Layer-1 to the average leakage current of modules on Layer-2, are shown in  Figure~\ref{fig:OuterPixelTimeRatio} for LHC Run~2. These ratios are, as predicted, fairly constant as a function of integrated luminosity. Once again, some dates corresponding to extended periods when the LHC beams were off are displayed with grey vertical lines. Also shown in  Figure~\ref{fig:OuterPixelTimeRatio} are the ratios of the unscaled Hamburg Model predictions for LHC Run~2. The vertical axis is proportional to the ratio of the applied fluences. The fluence of one layer relative to other layers is well predicted without the need for scale factors.
 
\clearpage
 
\subsection{Strip detector and disks}

The ATLAS SCT consists of 4088 modules of silicon-strip
detectors: 2112 in the barrel region, and 988 per endcap region, EC-A or EC-C. The barrel modules
are mounted on four cylindrical supports and named as Barrels 3 to 6 (B3 to B6 for short). Each endcap has nine disks (labelled 1 to 9), each consisting of up to three rings of modules. The rings, in order of increasing radius, are named \textit{Inner}, \textit{Middle} and
\textit{Outer}.
 
\subsubsection{Temperature measurement}
 
Each barrel module consists of four rectangular silicon-strip sensors. Two sensors on each side are
daisy-chained together. Two sides of identical pairs are glued back-to-back with a 380~$\mu$m thick
anisotropic thermal pyrolytic graphite (TPG) substrate in between.
 
The endcap region has four module types and five different trapezoid-shaped sensors. Modules in
the \textit{Outer} and \textit{Middle} rings consist of two daisy-chained sensors on each side, whereas those in the
\textit{Inner} ring have one sensor per side. A TPG spine sandwiched between two sensor sides conducts heat
away from the sensors to the cooling block contacts. The modules in the \textit{Outer} and \textit{Middle} are supported and
cooled by their contacts with two cooling blocks; the main block is shared between the hybrid and the
spine, while the far block cools only the spine. Inner modules are cooled only through the main
block.
 
SCT modules are cooled using the inner detector's evaporative $\text{C}_{3}\text{F}_{8}$ cooling system \cite{Attree_2008}. Each loop cools 48~(up to
33) barrel (endcap) modules. A total of 44 and 72 cooling loops are operated simultaneously for the barrel and endcap regions, respectively. The
cooling-pipe temperatures were set to about $-12$\degC, $-7$\degC, $-13$\degC and $-10$\degC in B3--B5, B6, EC-A and
EC-C, respectively. The cooling temperatures for B6 were set higher because of a failure in resistive
pad heaters on the thermal enclosure cylinders at the SCT--TRT interface. There is a 3\degC difference in the
cooling temperature settings of EC-A and EC-C due in part to different assembly sites. The initial cooling temperature settings were kept
from 2009 to 2018 except for 6~months during 2015 to avoid condensation in the SCT volumes.
 
The sensor temperature $T_\text{sensor}$ of each module is deduced from a hybrid-board temperature $T_\text{hybrid}$
measured by NTC thermistors mounted on the hybrid circuit board. A temperature offset between
$T_\text{hybrid}$ and $T_\text{sensor}$ strongly depends on the mechanical and thermal structure. It also differs module-by-module due to slightly different thermal resistances of the hybrid-circuit, sensor and cooling
pipe. The temperature offset is determined by measuring the hybrid temperature and the leakage current of the sensors with and
without power applied to the hybrid low voltage (LV). The hybrid temperature and the leakage current data are collected at several cooling temperature settings. The
temperature differences are extracted by an interpolation of leakage currents vs $T_\text{hybrid}$ curves. The offsets are similar to predictions from the thermal
finite-element method. The individual differences are 3--5\degC and 12--20\degC in barrel and endcap
modules, respectively. The offsets of the B6 modules are not measured because the TRT would become too cold at the SCT--TRT interface if the LV power were not supplied to the B6 modules.
 
During long shutdowns in winters and LS1, as well as occasional power-cut cases, $T_\text{hybrid}$ data are
not available. In most cases, however, the temperature monitors of the evaporative cooling system
were active. Since there was no heat generation in the SCT volume, the temperature of the closest
cooling pipe could be used as an estimate of $T_\text{sensor}$, which occasionally reached as high as 20\degC.
 
The precision of the $T_\text{hybrid}$ measurement is determined by using temperature data (when all HVs and LVs are off) from the two thermistors
mounted on each side of a barrel module. These readings
agree well with a root-mean-square (RMS) uncertainty of 0.27\degC. A major uncertainty in $T_\text{hybrid}$ comes from the smallest digitization unit
of 0.33\degC used in the conversion from thermistor resistance to temperature. However, module-by-module fluctuations in the leakage currents cause larger spreads and thus such digitization effects smear
out once the mean of  \textgreater 30 modules belonging to the same module group is taken.
 
\subsubsection{Time evolution of leakage current}
All voltages and currents of the HV power supplies, as well as $T_\text{hybrid}$, were continuously monitored
and stored in a database called the Detector Control System Data Viewer~\cite{Tsarouchas:2026327}. For the leakage current of each module, typically 150
data points of leakage current are recorded every hour because of the presence of small ripples of about 0.2\%.
During a physics run lasting several hours, as the instantaneous luminosity goes down, the HV
current drops by 0.2--2\% (values in late 2018, none in Run~1) due to sensor self-heating. A simple
time-weighted average during each physics run is taken for the present study.
 
Another way to get the leakage current data is to read the leakage current mean values recorded in the gain calibration runs. In-beam leakage current averages are consistent with the mean values from nearby calibration run values within 1\%.
 
It should be noted that the leakage current depends on the HV applied to the sensor. Well above the full
depletion voltage, the leakage current increases by several percent per 100~V. In addition, due to the
filter resistance of about 12~k$\Omega$ in the HV supply lines, the true voltage on the sensor could be 20~V
less in the worst case for B3 modules in late 2018, introducing up to 1\% additional uncertainty in
the leakage current measurement.

\begin{figure}[h!]
\centering
\includegraphics[width=0.45\textwidth]{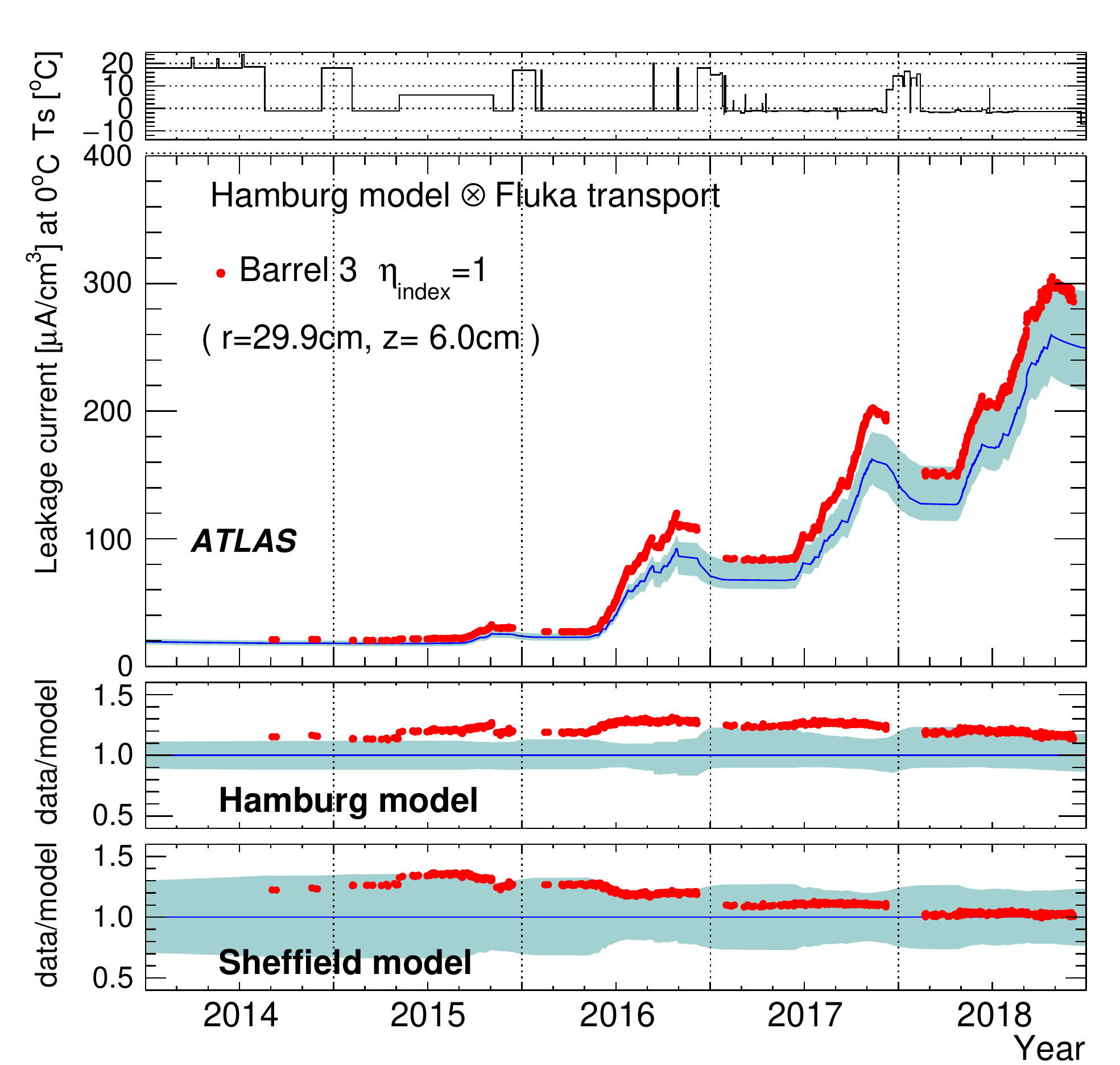}
\includegraphics[width=0.45\textwidth]{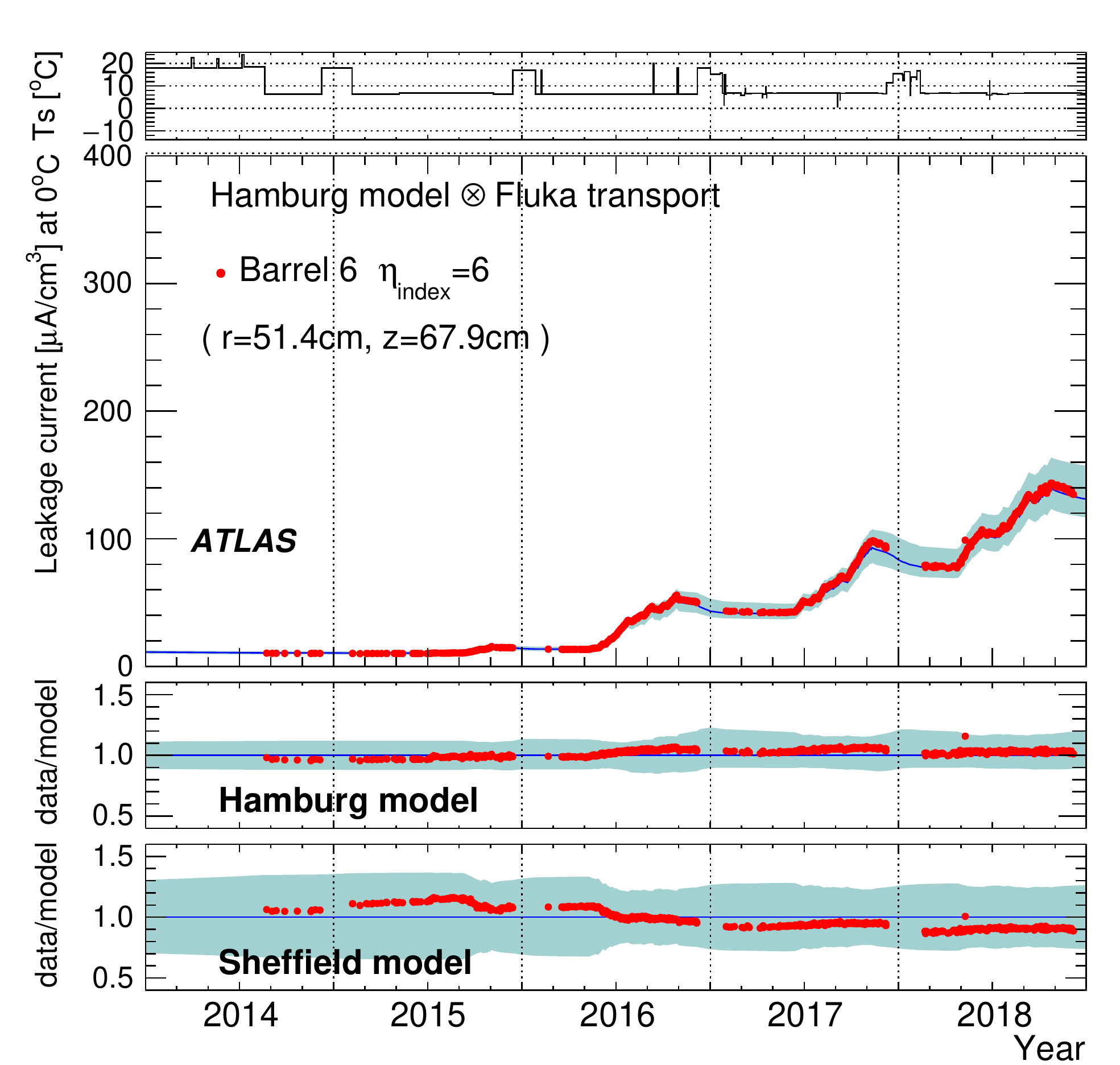}
\includegraphics[width=0.45\textwidth]{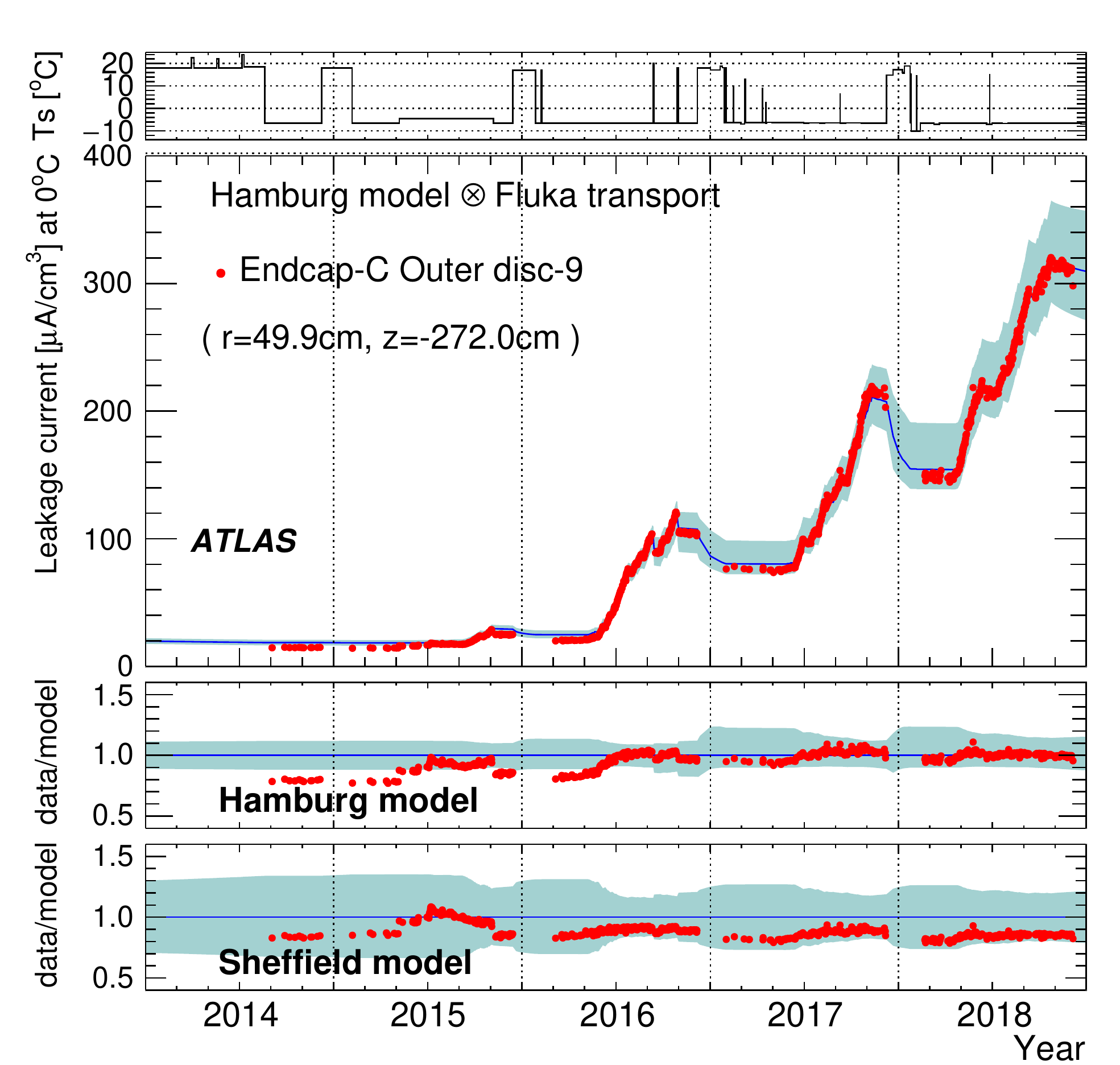}
\includegraphics[width=0.45\textwidth]{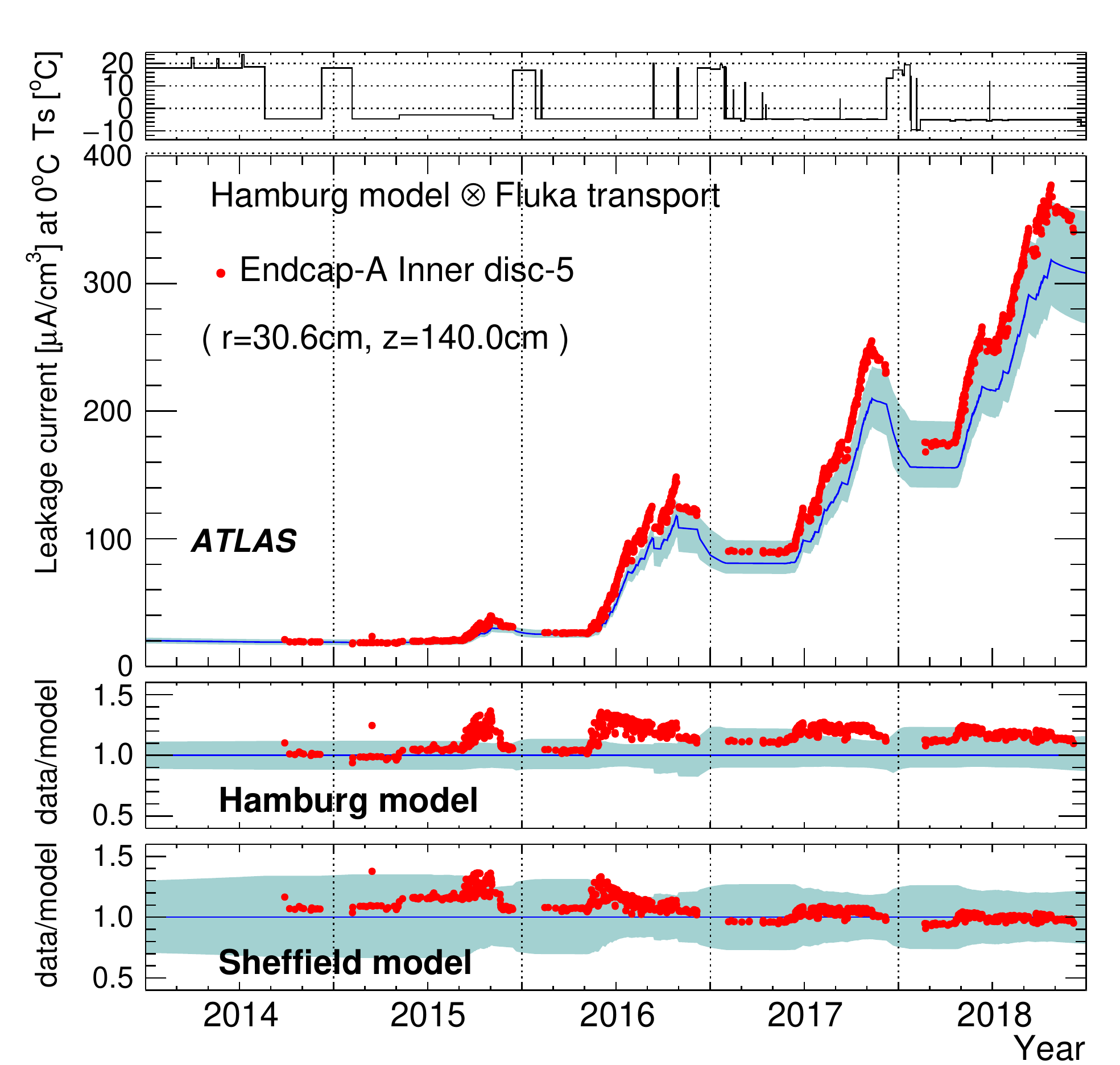}
\caption{
Typical time evolution of normalized leakage currents in the four SCT modules groups, one
each from B3, B6, EC-C and EC-A regions. Top plots show histories of deduced sensor temperatures,
while the second plots are leakage current data and Hamburg Model predictions. Each data point represents an average of time-weighted leakage current
means over 30--56 modules in a single physics or calibration run.  The two bottom plots show the ratios of data to predictions from the Hamburg and Sheffield models, using the same conversion factors as in \textsc{Fluka} transport simulations.  Coloured bands show 1$\sigma$ uncertainties of the model predictions.
}
\label{fig:sct1}
\end{figure}

Figure~\ref{fig:sct1} shows the time evolution of the leakage current during the Run~2 period including winter shutdowns for
four representative module groups, two each from barrel and endcap regions. The top plots show the
histories of the estimated sensor temperatures $T_\text{sensor}$, which were around $-1$\degC, $+5$\degC and $-7$\degC for B3, B6 and both
EC-A and EC-C, respectively, during the running periods, and were \textgreater 15\degC during LS1 and the winter shutdowns.
In the 2015 beam time, all modules except those in B6 were set warmer to avoid a condensation risk
in the SCT volume.  The HV is adjusted to always be above the prediction for full depletion.  The second set of plots from the top in Figure~\ref{fig:sct1} display normalized leakage currents, and the third and fourth sets of plots display ratios of data to predictions from the Hamburg Model and the Sheffield Model\footnote{This is an alternative to the Hamburg Model that has been developed for the ATLAS strip detector.  While the Hamburg model uses an exponential combined with a logarithmic function, the Sheffield model is based on a sum of five exponentials.  The parameters of these exponentials were tuned to SCT-like modules prior to the start of the LHC.  While it has not yet been compared with pixel data from any experiment, this will be important for the future (see Section~\ref{sec:disc}). }~\cite{HARPER2002548,Harper:2759942}.    Uncertainties from the model predictions are shown with bands which are calculated by varying each parameter of the model by 1$\sigma$ and adding the resulting changes in
quadrature. Uncertainties in the temperature measurements (1\degC) and delivered luminosities (3.7\%) are
also included.
 
The leakage current anneals out by
20--30\% during each winter shutdown, as evident from the drop in the current during periods of no beam.
The
Sheffield Model systematically predicts 15\% more leakage current than the Hamburg Model. For
most of the cases, the ratios seldom change by more than 5\% during a given year, which is good
evidence for the leakage current being proportional to irradiation with good-enough short annealing
terms. Any year-by-year dependencies in the ratios may be due to the influence of insufficiently accurate estimates of $T_\text{sensor}$ during
shutdowns or model limitations or both on annealing effects.
 
\subsubsection{Lateral distribution of leakage current}
\label{sec:sct3}
 
\begin{figure}[h!]
\centering
\begin{tabular}{cc}
\includegraphics[width=0.47\columnwidth]{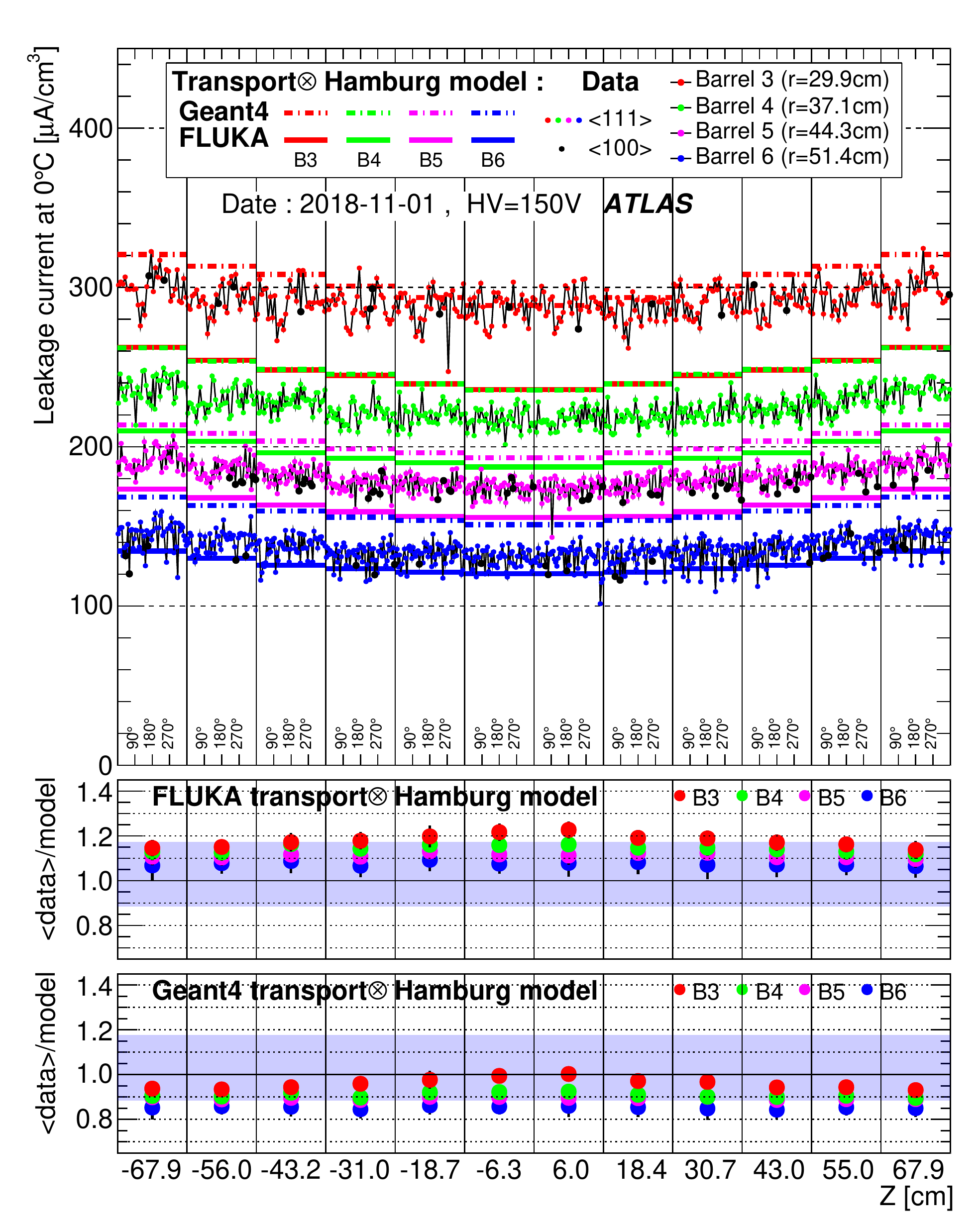} &
\includegraphics[width=0.47\columnwidth]{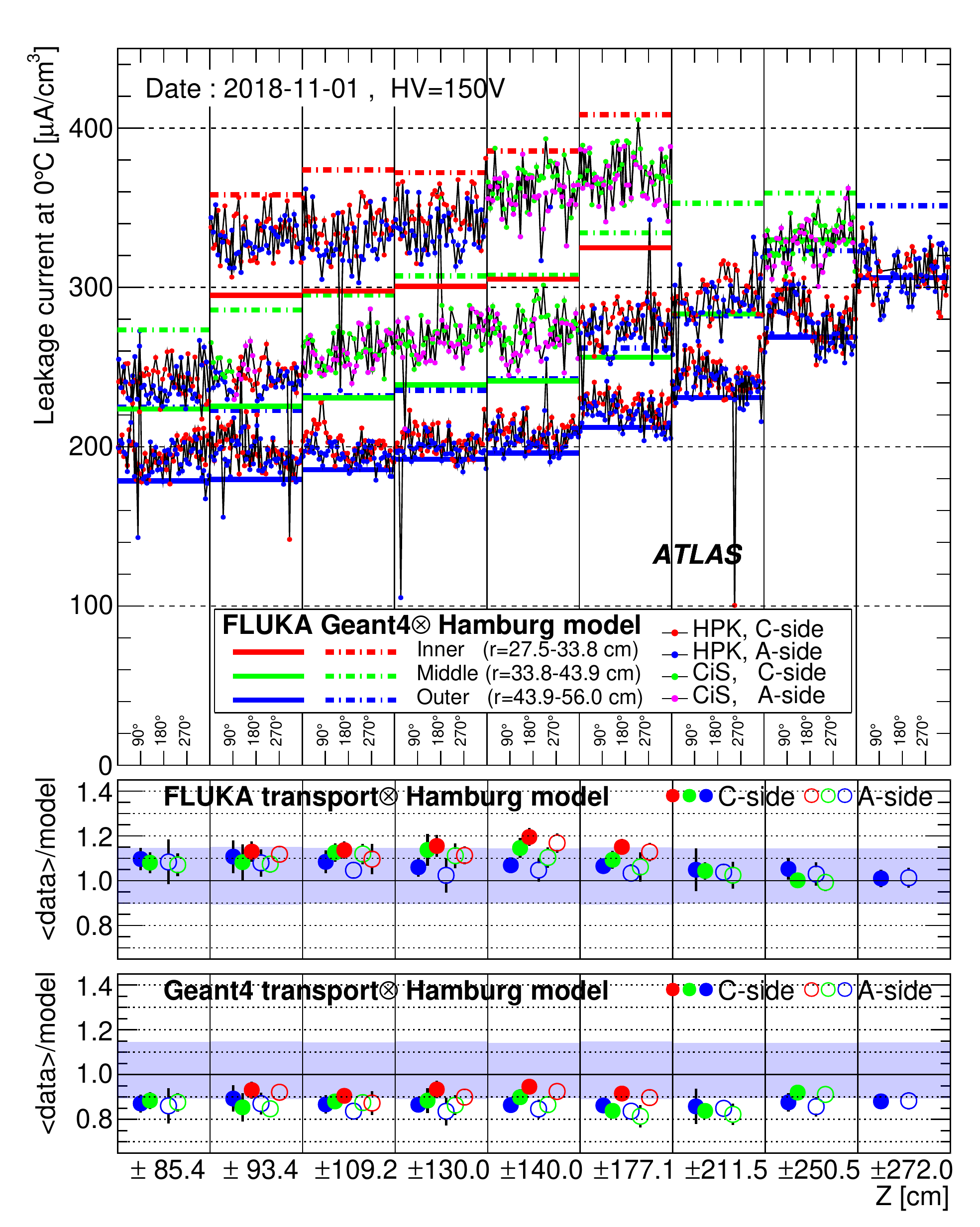} \\
(a) & (b) \\
\end{tabular}
\caption{ Leakage current measured at HV set to 150~V normalized to 0\degC per unit volume for all
(a) barrel modules and (b) endcap modules as of November 1, 2018. At the same $r$ and $z$ location,
modules with different $\phi$ indices are arranged horizontally from left ($\phi=0^\circ$) to right ($\phi=360^\circ$). Endcap
side-A and side-C as well as sensor manufacturers (Hamamatsu (HPK) and CiS) are plotted in
different colours. Horizontal solid/dot-dash bars indicate model predictions of the Hamburg Model
using the conversion factors by \textsc{Fluka} (solid) / \textsc{Geant} (dot-dash) transport simulations. In the two ratio
plots, mean and RMS values via Gaussian fits of modules belonging to the same group are plotted.
The model uncertainties are shown by blue bands.
}
\label{fig:sct2}
\end{figure}

Figures~\ref{fig:sct2}(a) and \ref{fig:sct2}(b) show leakage currents for all barrel and endcap modules, respectively, as of November
2018 with the applied HV set to 150~V. In these plots, modules with the same $r$ and $z$ locations
but different azimuthal angles are bundled side by side. Modules of EC-A and EC-C are coloured
differently, as are modules built with sensors from different manufacturers.
Permanently disabled
modules (42 in total out of about 2000) are not displayed in the plot. It can be seen that almost all modules in
the same group have quite similar leakage currents with a spread of about 3\%. Despite the
difference of 2--3\degC between the EC-A and EC-C cooling temperatures, the leakage currents agree well once they are normalized to 0\degC.
 
Additional differences between sensors across the detector did not result in appreciable variations among the leakage currents.  The majority of the modules are constructed from silicon wafers with crystal lattice orientation (Miller indices) $\langle 111\rangle$ while a small number of modules in the barrel use wafers with $\langle 100\rangle$ lattice orientation.  The barrel sensors and 75\% of the endcap sensors were supplied by Hamamatsu Photonics (HPK)\footnote{Hamamatsu Photonics Co. Ltd.,1126-1 Ichino-cho, Hamamastu, Shizuoka 431-3196, Japan.} while the remaining endcap sensors were supplied by CiS.\footnote{CiS Institut f\"{u}r Mikrosensorik gGmbH, Konrad-Zuse-Strasse 14, 99099 Erfurt, Germany.} Sensors supplied by the two manufacturers meet the same performance specifications, but differ in design and processing details \cite{Ahmad:2007zza}. The CiS sensors for EC inner modules were oxygen-enriched. No appreciable differences in leakage current were observed among sensors from different manufacturers (HPK vs CiS), or different crystal orientations ($\langle 111\rangle$ vs $\langle 100\rangle$) or standard/oxygen-enriched silicon materials.
One sees a clear and systematic trend of higher leakage currents in higher rapidity regions at all
radii covered by the SCT. In the barrel layers, the normalized leakage currents in near-centre modules
are about 3\% smaller than in edge modules at $|z|=68$~cm. This is in contrast to the
observation at the end of Run~1 when a slight excess was seen in the central B3 layer but a flat behavior was seen in other barrel layers \cite{IDET-2013-01}.
 
The observed gross trends are reproduced  fairly well by the Hamburg Model times the conversion
factors from \textsc{Fluka} or \textsc{Geant} transport simulations. In general, leakage current predictions
are systematically 10--20\% higher in the \textsc{Geant} case. The ratios in modules near the centre of B3 are 10\%
higher than in those at the edges, but the difference is less in the B6 layer, indicating an additional fluence
component close to the interaction point. In the endcap regions, however, the data-to-model ratios are
fairly constant for all disks in the range $|z|=85$ to 270~cm although their leakage currents being
different by up to 50\%.
 
\clearpage
 
\section{Leakage current and fluence comparisons}
\label{sec:comparisons}
 
This section incorporates measurements and simulations for all of the silicon-based inner-detector subsystems described in the previous sections.  For leakage current comparisons, the measurements are presented from the end of Run 2.  The simulations are combined with the thermal and luminosity history of the various sensors to transform fluence predictions to leakage current predictions for all detector regions.  For fluence comparisons, the annealing models are used to fit a scale factor that normalizes the fluence in the simulations per detector region.  The nominal predicted fluence combined with the scale factor gives the measured fluence rate, $\Phi/L_\text{int}$ (from Eq.~\ref{eq:leakage_modified}).
 
\begin{figure}[!ht]
\centering
\includegraphics[width=0.8\textwidth]{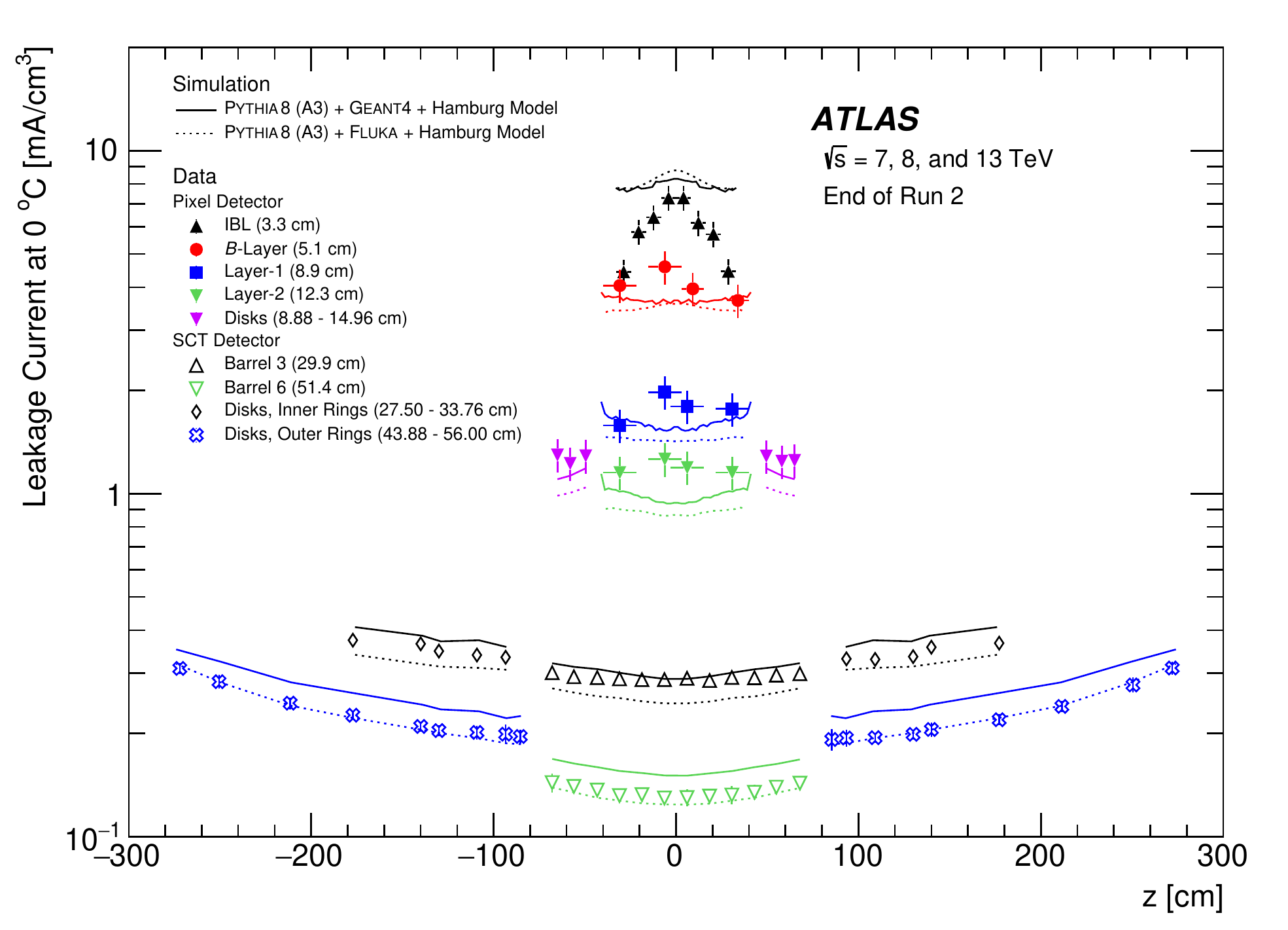}\\ \includegraphics[width=0.8\textwidth]{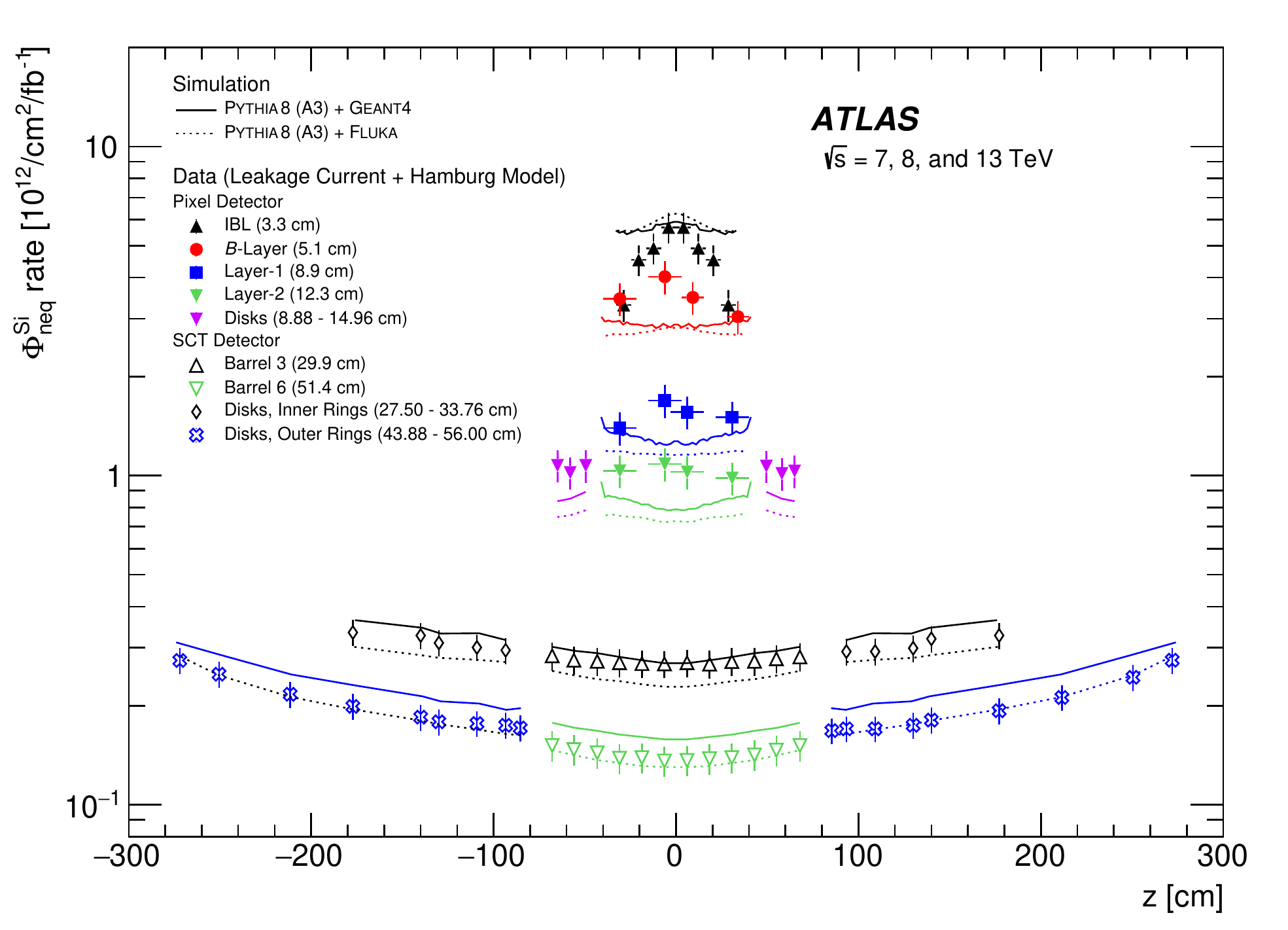}
\caption{The leakage current at the end of Run~2 (left) and the fluence rate (right) as a function of $z$ for the silicon-based parts of the ATLAS inner detector.   The predicted values are symmetric in $z$ by construction. Distances given in parentheses after layer names correspond to the radial positions of the sensors relative to the geometric centre of ATLAS.  For the IBL, the error bars are dominated by the residual dependence of the leakage current on the high voltage past full depletion; for the outer layers of the pixel detector, the uncertainty is dominated by a power supply uncertainty and uncertainties in the temperature and luminosity; for the SCT, the uncertainty is due to the sensor temperature, the luminosity, and the sensor thickness (for fluence) and the RMS spread across modules (leakage current).  Uncertainties in the silicon damage factors (relevant for the simulation and the Hamburg Model) are not included.}
\label{fig:combination}
\end{figure}

\begin{figure}[!ht]
\centering
\includegraphics[width=0.8\textwidth]{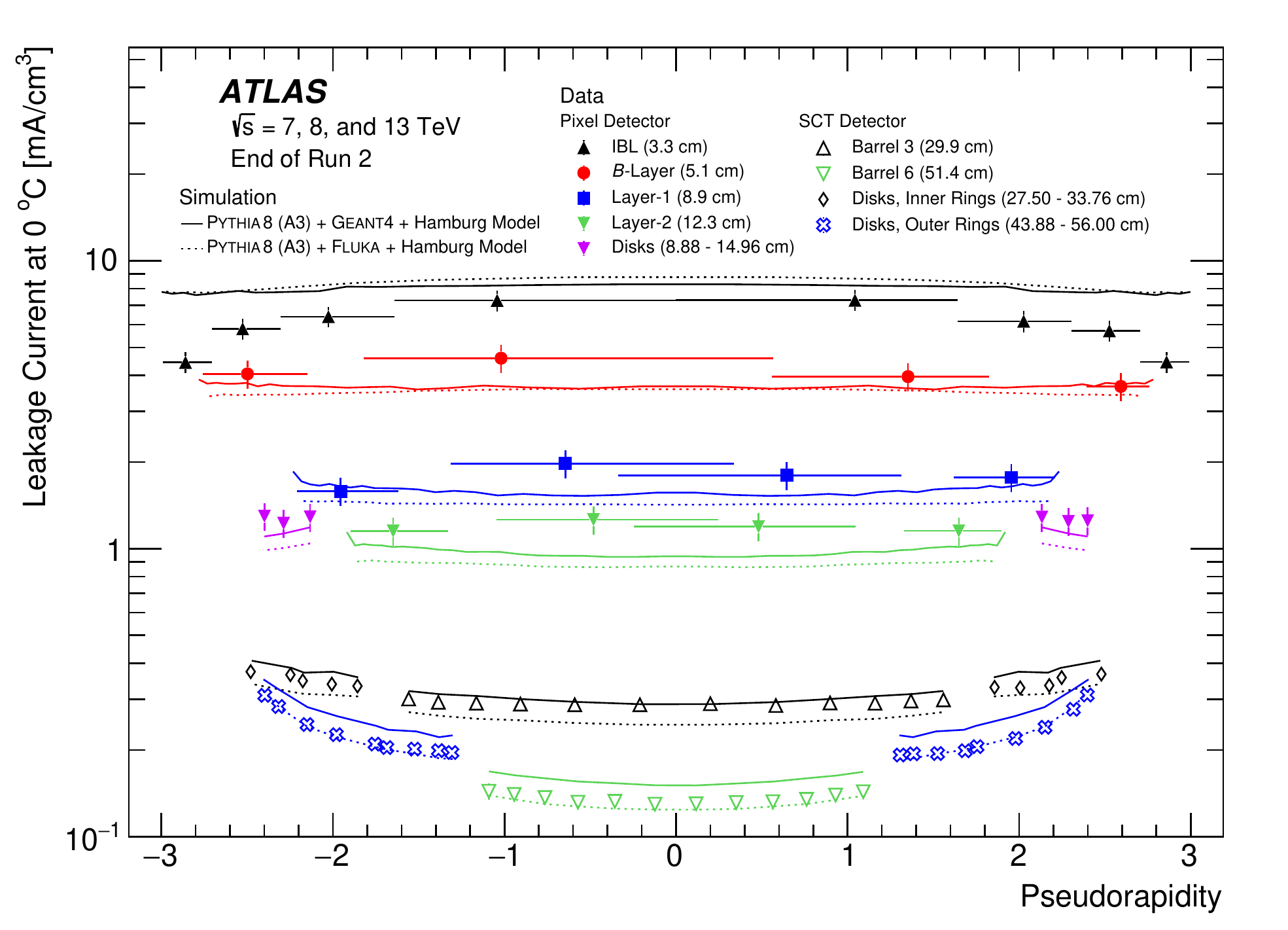}\\ \includegraphics[width=0.8\textwidth]{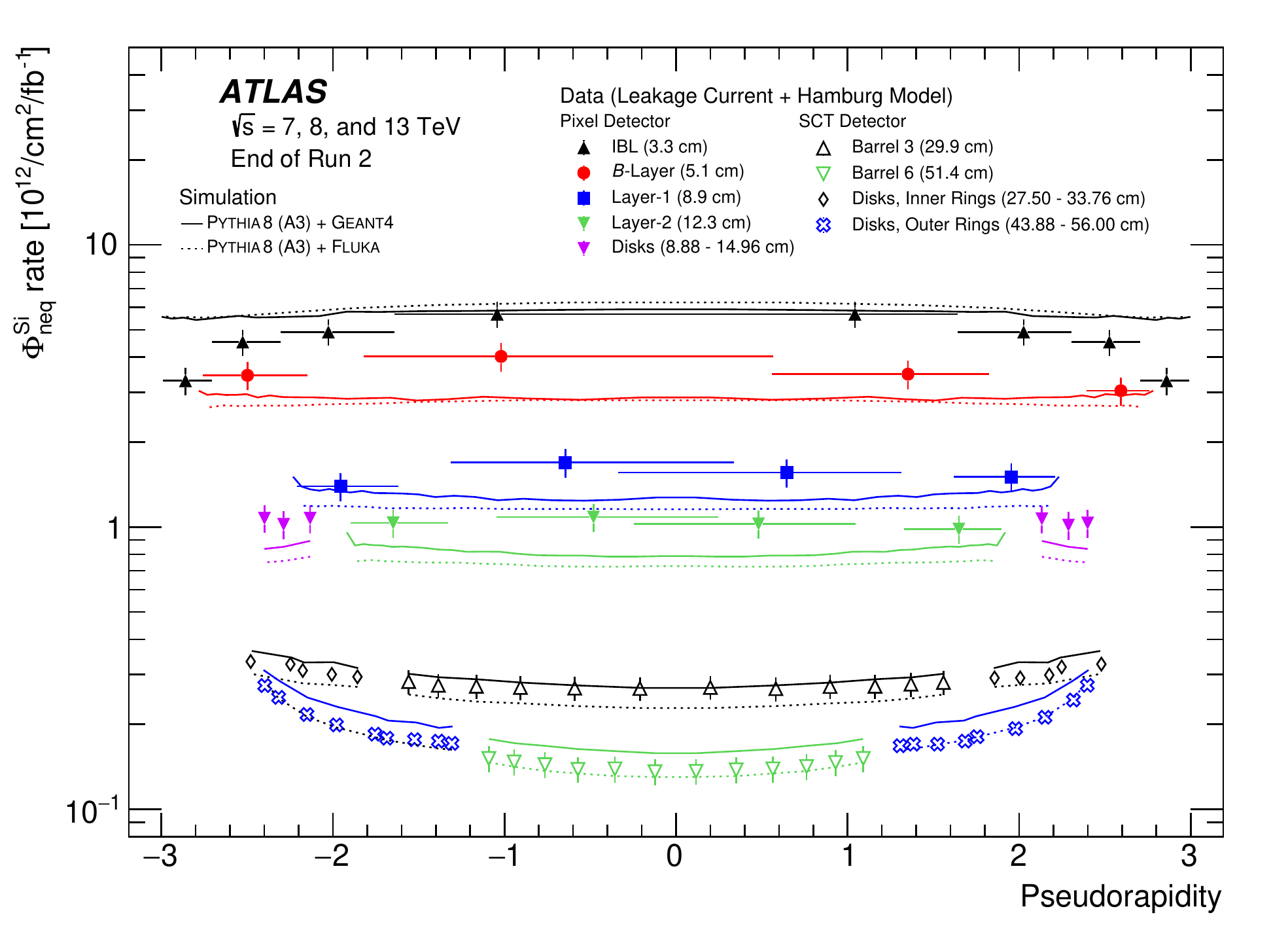}
\caption{The leakage current at the end of Run~2 (left) and the fluence rate (right) as a function of $\eta$ for the silicon-based parts of the ATLAS inner detector.   The predicted values are symmetric in $\eta$ by construction. Distances given in parentheses after layer names correspond to the radial positions of the sensors relative to the geometric centre of ATLAS.  For the IBL, the error bars are dominated by the residual dependence of the leakage current on the high voltage past full depletion; for the outer layers of the pixel detector, the uncertainty is dominated by a power supply uncertainty and uncertainties in the temperature and luminosity; for the SCT, the uncertainty is due to the sensor temperature, the luminosity, and the sensor thickness (for fluence) and the RMS spread across modules (leakage current).  Uncertainties in the silicon damage factors (relevant for the simulation and the Hamburg Model) are not included.}
\label{fig:combinationEta}
\end{figure}

Figure~\ref{fig:combination} presents a comparison of the measured leakage current and fluence rate for both the pixel and strip detectors.  The leakage current is the value at the end of Run~2 while the fluence rate is independent of time.\footnote{This is exactly true for the IBL, but for the other layers, a minor model dependence is introduced by using the relative total cross sections between collision energies (lower in Run~1 than in Run~2) to be able to measure a single value. About 85\% of the data come from Run~2, which further reduces the impact of this correction. }  The fluence rate prediction agrees well with the IBL data at $|z|=0$ and with all the SCT data.  In contrast, there is a much stronger $|z|$-dependence observed in the IBL data than is predicted (numerical values are presented in Table~\ref{tab:ScaleFactorsIBL}) and the overall fluence is significantly higher than predicted for the outer pixel layers.   Near the centre of the detector, the measured fluence is 30--50\% higher than predicted by the simulation.  For Layer-1 and Layer-2, the difference is 30--40\%.   For the IBL, the measured fluence at $|z|=30$~cm is about 50\% of the value at $|z|=0$.  These data are presented as a function of $\eta$ instead of $z$ in Figure~\ref{fig:combinationEta}.  The trends are exactly the same as in Figure~\ref{fig:combination}, but now the $|\eta| < 2.5$ acceptance of the silicon tracking detector is clear.  The fluence on the inner layers is mostly determined by the primary charged-pion flux, which is relatively constant as a function of $\eta$.  In contrast, a significant fraction of the fluence in the outer layers of the SCT is due to neutrons that are produced by interactions with material in the dense regions of the ATLAS calorimeters.
 
\begin{figure}[!ht]
\centering
\includegraphics[width=0.8\textwidth]{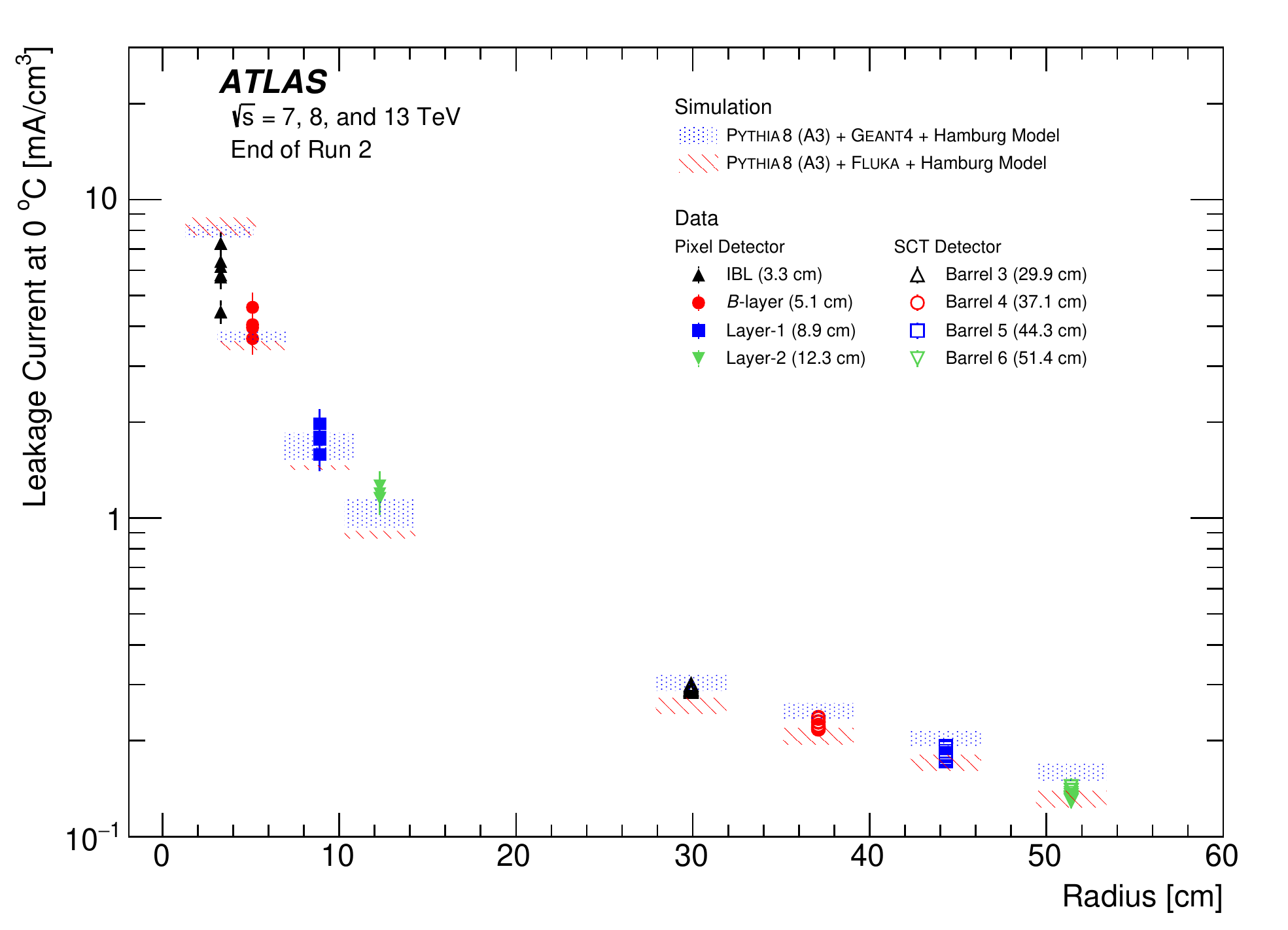}\\ \includegraphics[width=0.8\textwidth]{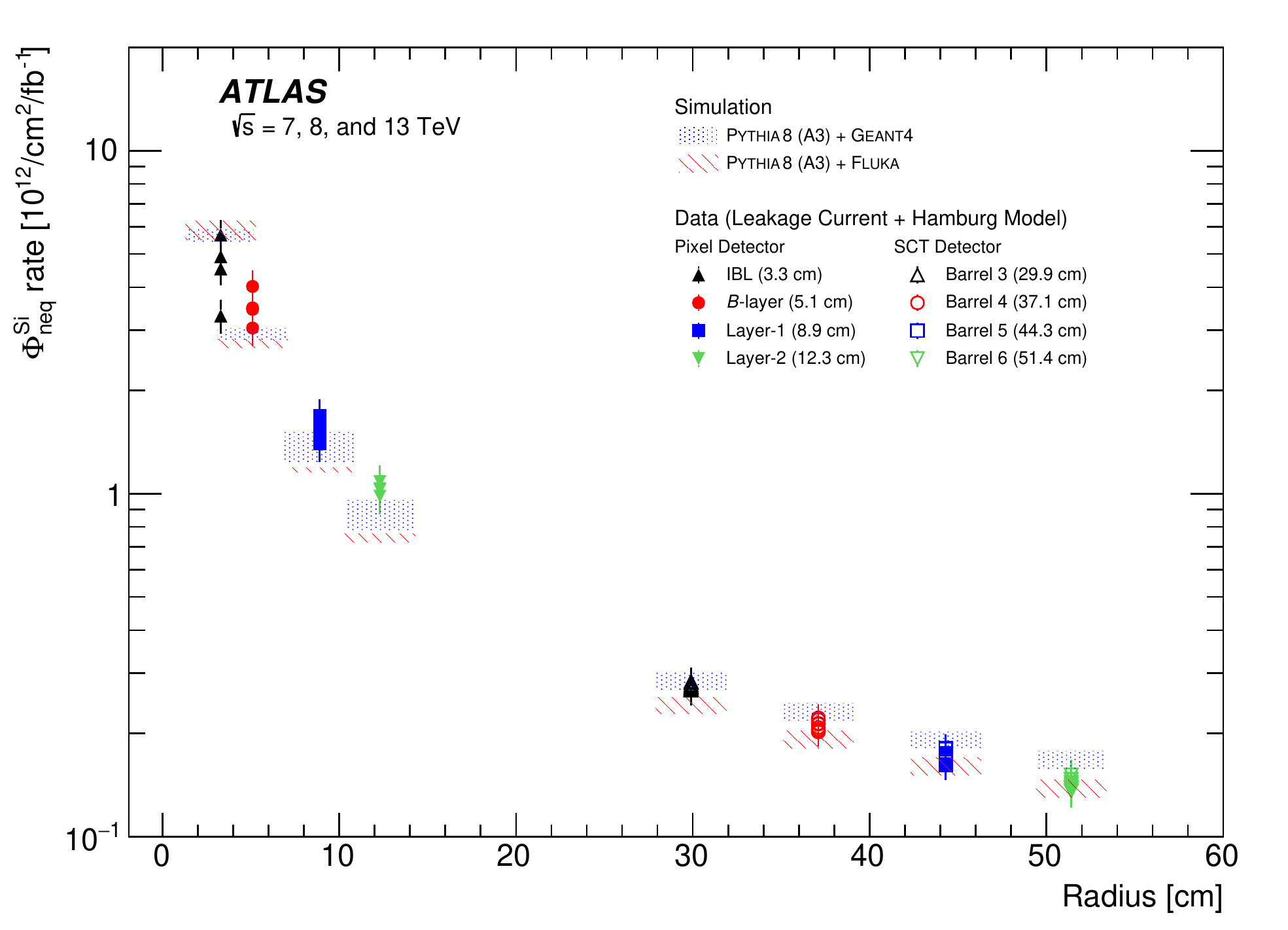}
\caption{The leakage current at the end of Run~2 (left) and the fluence rate (right) as a function of radius for the silicon-based parts of the ATLAS inner detector.  The length of the horizontal bands representing the simulation is chosen to aid the comparison with data -- the actual radial uncertainty from the finite size of the sensors is comparable to the marker sizes.  For the IBL, the error bars are dominated by the residual dependence of the leakage current on the high voltage past full depletion; for the outer layers of the pixel detector, the uncertainty is dominated by a power supply uncertainty and uncertainties in the temperature and luminosity; for the SCT, the uncertainty is due to the sensor temperature, the luminosity, and the sensor thickness (for fluence) and the RMS spread across modules (leakage current).  Uncertainties in the silicon damage factors (relevant for the simulation and the Hamburg Model) are not included.}
\label{fig:combinationRadius}
\end{figure}

The data are presented in a third way in Figure~\ref{fig:combinationRadius}, demonstrating the radial dependence of the measured leakage current and fluence.   The fluence drops off approximately as the square of the inverse radius, with deviations resulting from particles produced through interactions with the detector.   Beyond the pixel detector, the \textsc{Fluka} and \textsc{Geant}4 simulations bracket the measured values.
 
The ratio of the simulated values to the leakage current data, as shown in Figure~\ref{fig:combinationRadius}, is presented in Table~\ref{tab:ScaleFactors}. The values in this table  provide concrete input for further studies of the radiation environment at the LHC.
 
\begin{table}[!htb]
\caption{
IBL scale factors as a function of $z$, depicted in Figure~\ref{fig:combination}.   The measurements are consistent between $+z$ and $-z$ and the predictions are symmetric by construction, so the values are presented in bins of $|z|$.
}
\begin{center}
 
\begin{tabular}{r@{ }l@{ }c@{ }r@{ }lc}
\toprule \\[-9pt]
\multicolumn{5}{c}{$z$ Bin}       & Mean SF  \\
\midrule \\[-9pt]
$32$  &cm& $>$~$|z|$ $>$ & $24$ &cm  & 0.56   $\pm$ 0.06       \\
$24$  &cm& $>$~$|z|$ $>$& $16$ &cm  & 0.77     $\pm$  0.08         \\
$16$  &cm& $>$~$|z|$ $>$& $8$  &cm  & 0.84    $\pm$  0.09      \\
$8$  &cm& $>$~$|z|$ $>$& $0$   &cm  & 0.97     $\pm$  0.10       \\
\bottomrule
\end{tabular}
 
\end{center}
\label{tab:ScaleFactorsIBL}
\end{table}
 
\begin{table}[!htb]
\caption{
Mean, minimum, and maximum simulation-to-data ratios (scale factors, SF) for each barrel layer in the inner detector, as depicted in Figure~\ref{fig:combinationRadius}. The average uncertainty in the ratio is in the rightmost column.
}
\begin{center}
 
\begin{tabular}{ l l r r r r r}
\toprule \\[-9pt]
Detector & Layer & $r$ [cm] & Mean SF & Min.\ SF & Max.\ SF & SF uncert.\\
\midrule \\[-9pt]
\multirow{4}{*}{Pixel}  & IBL       &  3.30 & 0.78 & 0.56 & 0.97 & 0.08\\
& $B$-Layer &  5.10 & 1.28 & 1.11 & 1.47 & 0.15\\
& Layer-1   &  8.90 & 1.31 & 1.19 & 1.44 & 0.15\\
& Layer-2   & 12.30 & 1.39 & 1.32 & 1.46 & 0.16\\
\midrule \\[-9pt]
\multirow{4}{*}{SCT}    & Barrel 3   & 29.90 & 1.13 & 1.11 & 1.17 & 0.11\\
& Barrel 4   & 37.10 & 1.09 & 1.05 & 1.15 & 0.11\\
& Barrel 5   & 44.30 & 1.06 & 1.01 & 1.13 & 0.10\\
& Barrel 6   & 51.40 & 1.03 & 0.98 & 1.09 & 0.11\\
\bottomrule
\end{tabular}
 
\end{center}
\label{tab:ScaleFactors}
\end{table}
 
\subsection{Discussion}
\label{sec:disc}
 
The leakage current is a powerful probe of bulk damage in silicon caused by irradiation, and the results in the previous section indicate areas where simulations provide an excellent description of the data as well as other areas where there are significant deviations from observations.  The purpose of this section is to examine possible sources of the differences between data and simulation, including systematic effects for which there is currently no concrete uncertainty model.  Differences could be due to a variety of sources, affecting both the data measurements and simulation predictions:
 
\textbf{Method biases in the measurement}.  In theory, the current should rise, plateau, and then rise again as the high voltage is increased from zero up through breakdown. In practice, the current increases past full depletion for irradiated sensors so there is no unique high voltage at which to determine the leakage current.  Changes in the current past full depletion are typically small ($\mathcal{O}(10\%)$) but not negligible for the current level of measurement precision.   Furthermore, as remarked earlier, the current depends strongly on temperature.   This presents a challenge because the temperature of the sensors is often not known precisely. Temperature measurements are also complicated by bulk heat generation. This affects both the leakage current measurements and their interpretation because the measured temperature is used as an input for leakage current normalization and leakage current predictions.  A coupled complication is that the effective band-gap energy $E_\text{eff}$ may not be constant, as noted in Section~\ref{sec:eff}.  Additional measurements with other observables such as the Lorentz angle, depletion voltage, charge collection efficiency, etc.\ may add valuable information to confirm the trends observed with the leakage current.
 
\textbf{Physics modelling for the outgoing particle spectra}.  The input to radiation damage model predictions is the type and energy of particles produced by the primary proton--proton collisions.  The physics of soft quantum chromodynamics governs the majority of particles produced and these dynamics are not well understood.  A variety of models exist, as do measurements of the total inelastic cross section~\cite{Aaboud:2016mmw,Sirunyan:2018nqx,Aaij:2018okq} and minimum-bias / underlying-event spectra~\cite{Aaboud:2016itf,Aaboud:2017fwp,Khachatryan:2015jna}.  Systematic studies of model uncertainties and data/simulation differences may provide insight into deviations observed in the leakage current measurement.
 
\textbf{Transport models}.  Most of the significant data/prediction differences are common to both \textsc{Fluka} and \textsc{Geant}4. An important difference between the two transport models is that \textsc{Geant}4 includes a more detailed description of the ATLAS detector geometry. Improvements in the description of the inner-detector material may mitigate discrepancies, but the material is known precisely from studies of secondary interactions~\cite{Aaboud:2017pjd}.
 
\textbf{Damage factors}.   The largest single (largely unknown) uncertainty comes from the damage factors.  These factors have been tabulated by the RD50 Collaboration~\cite{273529,,HUHTINEN1993580,KONOBEYEV1992117,osti_10115441,art:Moll} and are based on a combination of measurements and simulations.  A key challenge is that there are not many beam facilities with monochromatic beams of hadrons in the relevant energy range.  Furthermore, the damage from neutrons changes rapidly near 1~\MeV\ so there is a significant uncertainty when converting from the damage from pions to the damage from neutrons.  Improving the precision of these factors is challenging, but a first important step would be to estimate their uncertainty.
 
\textbf{Annealing models}.   The Hamburg Model is the community standard for leakage current modelling.  Aside from the overall luminosity-to-fluence conversion, it has achieved  excellent precision over the full lifetime of the LHC.
However, there may be early indications from the pixel data that there are systematic differences between the model and data and these may grow to become significant in the future.
Alternative models are available, such as the Sheffield Model~\cite{HARPER2002548,Harper:2759942}.  Comparisons with this model as well as updating/tuning the Hamburg Model (including how to model periods of non-constant temperature) may be necessary for describing the leakage currents when including Run~3 and the high-luminosity phase of the LHC.
 
Despite these challenges, the leakage current is an important tool that will continue to improve in its descriptive and predictive power with additional studies on various fronts described above.
 
\FloatBarrier
 
\section{Conclusions and outlook}
\label{sec:conclusions}
 
This paper presents a measurement of the sensor leakage current for all the silicon detectors in the ATLAS tracking detector.  Over time and position within the detector, the existing models provide a reasonable description of the data.  Two significant discrepancies have been observed: there is a stronger $|z|$ dependence in the innermost layers than predicted by simulations and the overall fluence appears to be up to 50\% higher than in simulation for the intermediate layers between 5~cm and 15~cm from the collision point.  Overall, the fluence delivered to the innermost layer of the pixel detector is about $10^{15}$ 1~\MeV\ $n_\text{eq}/\text{cm}^2$ while the innermost portion of the SCT detector has experienced about $6\times 10^{13}$ 1~\MeV\ $n_\text{eq}/\text{cm}^2$.  The damage caused by these fluences has degraded the detector performance, but continued monitoring and modelling will support development of operational and offline analysis strategies to mitigate the impact on the physics output of the experiment.  Sensors designed for the high-luminosity phase of the LHC will need to cope with about an order of magnitude more fluence, and the investigations presented here will provide valuable input to preparations for those data-taking conditions in the near future.
\FloatBarrier
\section*{Acknowledgements}

We thank CERN for the very successful operation of the LHC, as well as the
support staff from our institutions without whom ATLAS could not be
operated efficiently.
 
We acknowledge the support of ANPCyT, Argentina; YerPhI, Armenia; ARC, Australia; BMWFW and FWF, Austria; ANAS, Azerbaijan; SSTC, Belarus; CNPq and FAPESP, Brazil; NSERC, NRC and CFI, Canada; CERN; ANID, Chile; CAS, MOST and NSFC, China; COLCIENCIAS, Colombia; MSMT CR, MPO CR and VSC CR, Czech Republic; DNRF and DNSRC, Denmark; IN2P3-CNRS and CEA-DRF/IRFU, France; SRNSFG, Georgia; BMBF, HGF and MPG, Germany; GSRT, Greece; RGC and Hong Kong SAR, China; ISF and Benoziyo Center, Israel; INFN, Italy; MEXT and JSPS, Japan; CNRST, Morocco; NWO, Netherlands; RCN, Norway; MNiSW and NCN, Poland; FCT, Portugal; MNE/IFA, Romania; JINR; MES of Russia and NRC KI, Russian Federation; MESTD, Serbia; MSSR, Slovakia; ARRS and MIZ\v{S}, Slovenia; DST/NRF, South Africa; MICINN, Spain; SRC and Wallenberg Foundation, Sweden; SERI, SNSF and Cantons of Bern and Geneva, Switzerland; MOST, Taiwan; TAEK, Turkey; STFC, United Kingdom; DOE and NSF, United States of America. In addition, individual groups and members have received support from BCKDF, CANARIE, Compute Canada, CRC and IVADO, Canada; Beijing Municipal Science \& Technology Commission, China; COST, ERC, ERDF, Horizon 2020 and Marie Sk{\l}odowska-Curie Actions, European Union; Investissements d'Avenir Labex, Investissements d'Avenir Idex and ANR, France; DFG and AvH Foundation, Germany; Herakleitos, Thales and Aristeia programmes co-financed by EU-ESF and the Greek NSRF, Greece; BSF-NSF and GIF, Israel; La Caixa Banking Foundation, CERCA Programme Generalitat de Catalunya and PROMETEO and GenT Programmes Generalitat Valenciana, Spain; G\"{o}ran Gustafssons Stiftelse, Sweden; The Royal Society and Leverhulme Trust, United Kingdom.
 
The crucial computing support from all WLCG partners is acknowledged gratefully, in particular from CERN, the ATLAS Tier-1 facilities at TRIUMF (Canada), NDGF (Denmark, Norway, Sweden), CC-IN2P3 (France), KIT/GridKA (Germany), INFN-CNAF (Italy), NL-T1 (Netherlands), PIC (Spain), ASGC (Taiwan), RAL (UK) and BNL (USA), the Tier-2 facilities worldwide and large non-WLCG resource providers. Major contributors of computing resources are listed in Ref.~\cite{ATL-SOFT-PUB-2020-001}.

\clearpage
 
\printbibliography
 
\clearpage 
 
\begin{flushleft}
\hypersetup{urlcolor=black}
{\Large The ATLAS Collaboration}

\bigskip

\AtlasOrcid[0000-0002-6665-4934]{G.~Aad}$^\textrm{\scriptsize 101}$,    
\AtlasOrcid[0000-0002-5888-2734]{B.~Abbott}$^\textrm{\scriptsize 128}$,    
\AtlasOrcid[0000-0002-7248-3203]{D.C.~Abbott}$^\textrm{\scriptsize 102}$,    
\AtlasOrcid[0000-0002-2788-3822]{A.~Abed~Abud}$^\textrm{\scriptsize 36}$,    
\AtlasOrcid[0000-0002-1002-1652]{K.~Abeling}$^\textrm{\scriptsize 53}$,    
\AtlasOrcid[0000-0002-2987-4006]{D.K.~Abhayasinghe}$^\textrm{\scriptsize 93}$,    
\AtlasOrcid[0000-0002-8496-9294]{S.H.~Abidi}$^\textrm{\scriptsize 29}$,    
\AtlasOrcid[0000-0002-8279-9324]{O.S.~AbouZeid}$^\textrm{\scriptsize 40}$,    
\AtlasOrcid{N.L.~Abraham}$^\textrm{\scriptsize 156}$,    
\AtlasOrcid[0000-0001-5329-6640]{H.~Abramowicz}$^\textrm{\scriptsize 161}$
% ...
\bigskip
\\

% ...
$^{29}$Physics Department, Brookhaven National Laboratory, Upton NY; United States of America.\\
% ...
$^{36}$CERN, Geneva; Switzerland.\\
% ...
$^{40}$Niels Bohr Institute, University of Copenhagen, Copenhagen; Denmark.\\
% ...
$^{53}$II. Physikalisches Institut, Georg-August-Universit\"{a}t G\"ottingen, G\"ottingen; Germany.\\
% ...
$^{93}$Department of Physics, Royal Holloway University of London, Egham; United Kingdom.\\
% ...
$^{101}$CPPM, Aix-Marseille Universit\'e, CNRS/IN2P3, Marseille; France.\\
$^{102}$Department of Physics, University of Massachusetts, Amherst MA; United States of America.\\
% ...
$^{128}$Homer L. Dodge Department of Physics and Astronomy, University of Oklahoma, Norman OK; United States of America.\\
% ...
$^{156}$Department of Physics and Astronomy, University of Sussex, Brighton; United Kingdom.\\
% ...
$^{161}$Raymond and Beverly Sackler School of Physics and Astronomy, Tel Aviv University, Tel Aviv; Israel.\\
% ...

\end{flushleft}


\end{document}